\pdfoutput=1
\documentclass[runningheads]{llncs}

\usepackage{mysty}
\usepackage[inline]{enumitem}

\usepackage{changes}

\definechangesauthor[name={AA}, color=blue]{AA}
\definechangesauthor[name={AA}, color=blue]{AP}

\title{Active Learning Techniques for Pomset Recognizers}

\author{Adrien Pommellet \inst{1}\orcidlink{0000-0002-1825-0097} \and
	Amazigh Amrane\inst{1}\orcidlink{0009-0008-7577-6715} \and
	Edgar Delaporte \inst{1}\and Geoffroy Du Prey\inst{1} \and Oscar Peyron \inst{1}\inst{2}}

\institute{LRE, EPITA, Le Kremlin-Bicêtre, France \and  
	École Polytechnique, France}
\authorrunning{A.~Pommellet, A.~Amrane, E.~Delaporte, G.~Du Prey, O.~Peyron}

\begin{document}

	\maketitle

	\begin{abstract}
		Series-parallel pomsets are a promising mathematical formalism for concurrent programs, as they can be recognized by simple algebraic structures known as pomset recognizers.
		Active learning consists in inferring a formal model of a system by interactively probing its behavior through queries to a Minimally Adequate Teacher (MAT).
		We improve existing learning algorithms for pomset recognizers by
		\begin{enumerate*}
			\item designing a new counterexample analysis procedure that is in the best case scenario exponentially more efficient than existing techniques,
			\item introducing and implementing a new algorithm $\PLlambda$ that extends the state-of-the-art $\Llambda$ algorithm to pomset recognizers, minimizing the impact of exceedingly verbose counterexamples and removing redundant queries, and
			\item designing a suitable finite test suite that ensures equivalence between two pomset recognizers by adapting the well-known $W$-method.
		\end{enumerate*}
		
	\end{abstract}

	\section{Introduction}
	
	Finite state automata are a straightforward model for terminating sequential systems.
	Runs are described by a \emph{total} order relation: an execution is an ordered, linear sequence of events.
	But concurrent programs require richer structures.
	
	\begin{center}
		\begin{minipage}{0.56\linewidth}
			\setlength{\parindent}{15pt}
			Indeed, two threads may be acting in parallel, neither of them preceding nor following the other.
			In this case, runs may be modelled using a \emph{partial} order: concurrent events cannot be relatively ordered, but sequential ones can.
			\emph{Series-parallel partially ordered multisets}~\cite{Pratt86pomsets} (SP pomsets) offer a convenient linear description of such executions based on sequential and parallel composition of letters.
		\end{minipage}
		\hfill
		\begin{minipage}{0.40\linewidth}
			\centering
			\begin{figure}[H]
				\centering
				\begin{tikzpicture}[shorten >=5pt, scale=0.9]
					\node (A) {$a$};
					\node[above right of=A] (B1) {$b$};
					\node[below right of=A] (B2) {$b$};
					\node[below right of=B1] (C) {$c$};
					\node[above right of=C] (D1) {$b$};
					\node[below right of=C] (D2) {$b$};
					\node[right of=D1] (E1) {$a$};
					\node[right of=D2] (E2) {$b$};
					
					\path[shorten <= 5pt, line width=0.1pt]
					(A.center) edge node {} (B1.center)
					(A.center) edge node {} (B2.center)
					(B1.center) edge node {} (C.center)
					(B2.center) edge node {} (C.center)
					(C.center) edge node {} (D1.center)
					(C.center) edge node {} (D2.center)
					(D1.center) edge node {} (E1.center)
					(D2.center) edge node {} (E2.center)
					;
				\end{tikzpicture}
				\caption{The series-parallel pomset $a (b \parallel b) c (ba \parallel bb)$.}
				\label{fig:sp_pomset}
			\end{figure}
		\end{minipage}
	\end{center}
	
	As an example, Figure~\ref{fig:sp_pomset} displays the Hasse diagram of the SP pomset $a (b \parallel b) c (ba \parallel bb)$.
	This model is in some cases exponentially more succinct than words:
	a single pomset $a_1 \parallel \ldots \parallel a_n$ describes the interleaving semantics of $n$ parallel threads and subsumes the $n!$ possible linearized traces of the interwoven threads.
	
	Languages of pomsets can be efficiently recognized by a special class of deterministic bottom-up tree automata known as \emph{pomset recognizers}, which are derived from an algebraic framework studied by Lodaya and Weil \cite{LW00:sp} called \emph{bimonoids} \cite{bloom1996free}: sets equipped with two internal operations, one associative and the other associative and commutative, with a neutral element for each operation.
	
	\emph{Active learning} (AL) consists in inferring a formal model of a black-box system that can be dynamically queried.
	Under \emph{the Minimally Adequate Teacher} (MAT) framework, interactions with the black-box system are twofold:
	\emph{membership queries} (MQs) to ask whether a given trace can be generated by the system, and \emph{equivalence queries} (EQs) to determine whether a given formal model (known as the \emph{hypothesis}) accurately represents all executions of the system, returning a counterexample that refines the hypothesis if the answer is negative.
	
	One of the earliest AL algorithms is Angluin's $\Lstar$~\cite{DBLP:journals/iandc/Angluin87} for rational languages.
	In a FoSSaCS 2021 article, van Heerdt et al.~\cite{DBLP:conf/fossacs/HeerdtKR021} applied $\Lstar$ to the class of recognizable pomset languages (we denote this extension $\PLstar$).
	Over the years, various improvements have been brought to the original $\Lstar$:
	\begin{itemize}
		\item New algorithms such as TTT~\cite{isberner14ttt}, $L^\#$~\cite{vaandrager22lsharp}, or $\Llambda$~\cite{DBLP:conf/birthday/HowarS22} have been shown to significantly reduce the number of MQs performed, e.g. through the use of redundancy-free discrimination trees~\cite{kearns94clt}.
		
		\item Since the counterexample's length $m$ returned by the MAT can be arbitrarily large, it may dominate the AL process.
		Rivest and Schapire~\cite{DBLP:journals/iandc/RivestS93} proposed an algorithm that refines the hypothesis using only $\bigo(\log(m))$ MQs.		
		
		\item Using EQs makes little practical sense as it assumes that the MAT knows the very formal model $M$ of the system being inferred;
		Chow~\cite{DBLP:journals/tse/Chow78} and Vasilevskii \cite{Vasilevskii1973} independently showed that, given a bound on $|M|-|H|$, a test suite of polynomial size w.r.t. $|H|$ can subsume equivalence between $M$ and $H$.
	\end{itemize}

	However, very few AL algorithms for intrinsically concurrent models have been designed so far.
	In order to advance automated reasoning about concurrent behaviours, we adapt and extend state-of-the-art techniques to recognizable pomset languages.
	Our new contributions are the following:
	\begin{description}
		\item[A new counterexample analysis algorithm.] Assume that a counterexample has a tree representation of depth $d$ that features $m$ nodes.
		Existing techniques infer a refinement in $\bigo(m)$ MQs. 
		We introduce a method that only requires $\bigo(d)$ queries instead (hence $\bigo(\log(m))$ for balanced trees).
		
		\item[Extending $\Llambda$~\cite{DBLP:conf/birthday/HowarS22} to recognizable pomset languages.] Algorithm $\Llambda$ is a state-of-the-art active learning algorithm for rational languages which has shown better empirical results than $L^*$.
		We design an expansion $\PLlambda$ to recognizable pomset languages that shares the same empirical behavior.
		
		\item[Adapting the $W$-method~\cite{DBLP:journals/tse/Chow78,Vasilevskii1973}.] We design a finite test suite that can conditionally replace EQs.
		
		\item[A \texttt{C++} tool.] We implemented and benchmarked our algorithms as well as van Heerdt et al.'s~\cite{DBLP:conf/fossacs/HeerdtKR021} $\PLstar$ (that, until now, had yet to be implemented).
	\end{description}
	
	Finally, it is worth noting that these algorithms can be applied as is to binary tree automata, associativity and commutativity constraints notwithstanding.

	\subsection{Related works}
	\label{sec:related}
	
%	Different classes of pomsets~\cite{DBLP:journals/iandc/FahrenbergJSZ22,VTL82:SPDigraphs} accepted by different automata~\cite{DBLP:conf/concur/FahrenbergJSZ22,lodaya01kleene} have been developed over the years, reflecting various communication models and interpretations of concurrency.	

	\paragraph{Pomset languages.}
	
	Automata over series-parallel pomsets, known as \emph{branching automata}, were first introduced by Lodaya and Weil~\cite{LW98:Algebra,LW00:sp}, as well as a generalization of regular expressions~\cite{lodaya98kleene,lodaya01kleene} to pomset languages.
	\emph{Pomset automata} were then introduced in~\cite{DBLP:journals/jlp/KappeBLSZ19};
	it was later shown in~\cite{DBLP:conf/fsttcs/Bedon21} that branching automata and pomset automata are effectively equivalent.
	From an algebraic perspective, Lodaya and Weil investigated the recognizability of languages of SP pomsets using \emph{bimonoids}. 
	They proved that recognizable languages adhere to a Myhill-Nerode-like theorem (hence, amenable to learning under a MAT) and are also recognized by branching automata, but that the converse does not hold.
	
	Alur et al.~\cite{alur2023robust} recently introduced synchronized series-parallel graphs, a variant of series-parallel pomsets that allows  ordered and unordered parallel compositions, and a corresponding class of automata and a logical characterization.
	They posit that these graphs could be used to model distributed data streams.
	
%	A translation from pomset recognizers to pomset automata has been provided~\cite{DBLP:conf/fossacs/HeerdtKR021} and
%	logical characterizations of branching automata and pomset recognizers have been established in \cite{DBLP:conf/mfcs/Bedon13} and \cite{DBLP:conf/dlt/Kuske01} respectively.
	
%	From an algebraic perspective, Lodaya and Weil defined \emph{SP-algebras} (from which pomset recognizers are derived): sets equipped with two inner products, one associative, the other associative and commutative.
	
	\paragraph{Active learning algorithms for finite state machines.}
	
	The original $L^*$ AL algorithm~\cite{DBLP:journals/iandc/Angluin87} maintains a table structure called an \emph{observation table} that it fills by calling MQs.
	\texttt{TTT}~\cite{isberner14ttt} instead features a \emph{discrimination tree} that has been experimentally shown to reduce the number of MQs needed.
	$L^\#$~\cite{vaandrager22lsharp} operates directly on a prefix tree that stores MQs and tries to establish apartness, a constructive form of non-equivalence.
	$\Llambda$~\cite{DBLP:conf/birthday/HowarS22} relies on partition refinement;
	its peculiarity is not adding substrings of counterexamples to its data structures.

	The last three algorithms have all been shown to be competitive and a net improvement over $L^*$.
	We chose to adapt $\Llambda$ due to its generic, unifying framework that is not intrinsically tied to finite words and rational languages.
%	$L^\#$'s prefix-closed trie structure is intrinsically tied to words, hence total orders, and \texttt{TTT} heavily relies on a complex automata-theoretic process to refine its discrimination tree that can seldom be generalized to pomset recognizers.

	\paragraph{Active learning algorithms for parallel models.}
	
	Pomset recognizers can be seen as a special case of deterministic bottom-up tree automata.
%	as a consequence, \textbf{(\romannumeral 1)} their equivalence is decidable. Thanks to \textbf{(\romannumeral 1)} and \textbf{(\romannumeral 2)}, it is possible to learn pomset recognizers, although the same algorithms do not apply to pomset automata because of \textbf{(\romannumeral 3)}. 
	Drewes and Högberg~\cite{DBLP:conf/dlt/DrewesH03} proposed an AL algorithm for tree automata, which later inspired the approach of van Heerdt et al.~\cite{DBLP:conf/fossacs/HeerdtKR021} for pomset recognizers.
	Whereas the latter directly infers new distinguishers from counterexamples, the former computes them inductively in a fashion similar to $\PLlambda$, although they failed to account for sharpness issues in the partition it maintains.
	
	Moreover, both Drewes and Högberg's algorithm as well as $\PLstar$ rely on a counterexample analysis procedure of linear complexity w.r.t. the
	input tree's number of nodes, whereas \FindEBP's complexity instead depends instead on its depth, hence, is logarithmic if the input tree is balanced.
	
	In~\cite{alduhaiby2020,DBLP:conf/concur/Henry0NS25,labbaf23,moerman2019pa,neele2023,samadi2025},
	\emph{compositional} learning algorithms for concurrent programs are introduced:
	the system under learning is decomposable as a parallel product of components individually modelled by Mealy or Moore machines.
	Rather than learning the entire system as a single composite product, each component is instead learnt in isolation.
	Depending on the algorithm considered, the alphabets of the various components may or may not be known beforehand.
	
	Compositional approaches are theoretically more frugal but less expressive than our pomset-based approach, as the resulting model learnt remains finite state.
	Indeed, given $w_1, w_2$ in the set $\SP$ of SP pomsets, we can inductively define a \emph{linearization} operator $\pi: \SP \to 2^{\Sigma^*}$ such that $\pi(a) = {a}$ for any letter $a \in \Sigma$, $\pi(w_1 \cdot w_2) = \pi(w_1) \cdot \pi(w_2)$ and $\pi(w_1 \parallel w_2) = \pi(w_1) \shuffle \pi(w_2)$, where $\shuffle$ is the \emph{shuffle} product on finite words.
	Given a partial order on a labelled set (i.e. a pomset), $\pi$ lists the total orders (i.e. finite words) on the same labelled set compatible with the aforementioned partial order.
	It allows us to directly compare the expressiveness of recognizable pomset languages and regular languages over finite words.
	
	Thanks to the pumping lemma, one can prove that there exists a recognizable pomset language $L \subseteq \SP$ such that $\pi(L) \subseteq \Sigma^*$ is not regular.
	Indeed, Example~\ref{ex:PR} defines one such language.
	Intuitively, its inductive definition allows for an arbitrary number of nested parallel branches;
	it models dynamic nested thread creation.

	\section{Preliminary Definitions}

	\subsection{Series-parallel pomsets}
	
	\begin{definition}[Series-parallel terms]
		It is the set $\terms$ over a finite \emph{alphabet} $\Sigma$ defined by the following grammar:
		\begin{equation*}
			t ::= {\varepsilon}~\mid~{a \in \Sigma}~\mid~{t \cdot t}~\mid~{t \parallel t}
		\end{equation*}
	\end{definition}
	
	\begin{center}
		\begin{minipage}{0.32\linewidth}
			\begin{figure}[H]
				\centering
				\begin{tikzpicture}[sibling distance=1.5cm, level distance=1cm]
					\node {$\parallel$}
					child {node (l) {$\cdot$}
						child {node (ll) {$b$}}
						child {node (lr) {$c$}}}
					child {node {$a$}};
				\end{tikzpicture}
				\caption{The term $bc \parallel a$.}
				\label{fig:synt_tree}
			\end{figure}
		\end{minipage}
		\hfill
		\begin{minipage}{0.65\linewidth}
			Series-parallel terms (from now on, \emph{terms}) are a particular class of full binary trees.
			The operators $\cdot$ and $\parallel$ are respectively called \emph{sequential composition} (or concatenation) and \emph{parallel composition}.
			Term $\varepsilon$ is called the \emph{empty term}. 
			A prefix traversal of a term yields a string known as its \emph{linear description}:
			syntactically, $\cdot$ is made implicit, $\cdot$ has priority over $\parallel$, and $\cdot$ and $\parallel$ are left associative.
			A term is shown in Figure~\ref{fig:synt_tree}.
		\end{minipage}
	\end{center}
	
	\begin{definition}[Series-parallel pomsets]
		The \emph{free algebra} of \emph{series-parallel pomsets}~\cite{bloom1996free} is the quotient space $\SP = {\terms} \qspace {\equiv_{\terms}}$ where $\equiv_{\terms}$ is the congruence relation such that $\varepsilon$ is a neutral element for $\cdot$ and $\parallel$, $\cdot$ and $\parallel$ are associative, and $\parallel$ is commutative w.r.t. $\equiv_{\terms}$.
	\end{definition}
		
	Intuitively, $a \cdot b$ stands for $b$ occurring after $a$, whereas $a \parallel b$ means that $a$ and $b$ are happening concurrently (hence the latter operator's commutativity).
	A series-parallel pomset (from now on, \emph{pomset}) models an execution trace of a concurrent program;
	a term is a description of this trace.
	
	We conventionally describe a pomset with the linear description of some representative:
	$a \parallel b = b \parallel a = a \cdot \varepsilon \parallel b$ all refer to the same pomset.
	Given $w \in \SP$, $\ST{w} \subseteq \terms$ stands for the set of terms representing the same pomset $w$ (that is, the pomset itself, were we to view it as an congruence class w.r.t. $\equiv_{\terms}$).
	Terms in $\ST{w}$ all share the same non-$\varepsilon$ leaves.
	
	We define the set $\SPplus = \SP \setminus \{\varepsilon\}$ of \emph{non-empty pomsets}.
	
	A term of $w$ is said to be \emph{minimal} if it is $\varepsilon$-free (or consists of $\varepsilon$ alone), of minimal depth w.r.t. $\ST{w}$, and its subterms (that is, its subtrees) are minimal too.
	The \emph{depth} $\depth{w}$ (resp. \emph{size} $|w|$) of $w$ is the minimum of the tree depth (resp. number of nodes) function on $\ST{w}$.
	Given $w_1, w_2 \in \SP$, $w_1$ is a \emph{subpomset} of $w_2$ if there exist $t_1 \in \ST{w_1}$ and $t_2 \in \ST{w_2}$ such that $t_2$ is a subterm of $t_1$.
	
	Historically, the word pomset \cite{Pratt86pomsets} stands for \emph{partially ordered multiset}: an isomorphism class of labelled, partially ordered sets.
	The \emph{parallel composition} $w_1 \parallel w_2$ (resp. \emph{sequential composition} $w_1 \cdot w_2$) of two pomsets $w_1$ and $w_2$ consists in assuming their elements are pairwise incomparable (resp. every element of $w_2$ is greater than every element of $w_1$).
	The set $\SP$ is the smallest set containing $\{\varepsilon\}$ and $\Sigma$, closed under sequential and parallel composition.
	It coincides with \emph{N-free} pomsets~\cite{VTL82:SPDigraphs}:
	there are no $x_1, x_2, x_3, x_4$ such that
	${x_1 < x_2}, {x_3 < x_2}, {x_3 < x_4}$ but $x_1$ and $x_4$ cannot be compared.

%	Historically, the word pomset \cite{Pratt86pomsets} stands for \emph{partially ordered multiset}: an isomorphism class of structures $(P, \leq, \ell)$, where $(P, \leq)$ is a partially ordered set (the carrier), $\ell : P \to \Sigma$, a labelling function, and isomorphisms, renamings of $P$'s elements that preserve both $\leq$ and $\ell$.
%	We always assume that the carrier sets of distinct pomsets are disjoint.
%	Given two pomsets $\mathfrak{A} = (A, {<_A}, \ell_A)$ and $w_2 = (B, {<_B}, \ell_B)$, their \emph{parallel composition} $\mathfrak{A} \parallel \mathfrak{B}$ is $(A \cup B, {<_A} \cup {<_B}, \ell_B \cup \ell_B)$:
%	the elements of $A$ and $B$ are pairwise incomparable.
%	Their \emph{sequential composition} $w_1 \cdot \mathfrak{B}$ is $(A \cup B, {<_A} \cup {<_B} \cup (A \times B), \ell_A \cup \ell_B)$:
%	every element of $B$ is greater than every element of $A$.
%	The set $\SP$ is the smallest set containing $\{\varepsilon\}$ and $\Sigma$, closed under sequential and parallel compositions.
%	It coincides with \emph{N-free} pomsets~\cite{VTL82:SPDigraphs}:
%	there are no $x_1, x_2, x_3, x_4$ such that
%	${x_1 < x_2}, {x_3 < x_2}, {x_3 < x_4}$ but $x_1$ and $x_4$ cannot be compared.

	\subsection{Pomset recognizers}
	
	\begin{definition}[Bimonoids~\cite{bloom1996free}]
		It is a tuple $(M, \odot, \obar, \neut)$ such that $M$ is a set equipped with two internal associative operations $\odot$ and $\obar$, $\obar$ being commutative as well, and a neutral element $\neut$ common to $\odot$ and $\obar$.
	\end{definition}
	
	\begin{definition}[Pomset recognizers]
		The tuple $\PR = (R, \odot, \obar, \neut, i, F)$ is said to be a \emph{pomset recognizer} (PR) over $\Sigma$ if $(R, \odot, \obar, \neut)$ is a finite bimonoid, $i \colon \Sigma \to R$, and $F \subseteq R$.
		The set $R$ (resp. $F$) is also called the set of \emph{states} (resp. \emph{accepting} or \emph{final} states) of $\PR$.
	\end{definition}
	
	We define the \emph{evaluation function} $\eval: \ST{\Sigma} \to R$ of $\PR$ inductively:
	$\eval(\varepsilon) = \neut$, if $a \in \Sigma$, $\eval(a) = i(a)$, $\eval(t_1 \cdot t_2) = \eval(t_1) \odot \eval(t_2)$ and $\eval(t_1 \parallel t_2) = \eval(t_1) \obar \eval(t_2)$.
	If $t_1 \equiv_{\terms} t_2$, then $\eval(t_1) = \eval(t_2)$:
	this property follows from the \emph{freeness} of $\SP$ \cite{bloom1996free} and from $\PR$ being a bimonoid.
	Thus, we can define $\eval: \SP \to R$ where, given $w \in \SP$, $\eval(w) = \eval(t)$ for any $t \in \ST{w}$.
	The \emph{language} of $\PR$ is the set $\Lang{\PR} = \{w \in \SP \mid \eval(w) \in F\}$.
	If $w \in \Lang{\PR}$, then $\PR$ \emph{accepts} $w$;
	we then write $\PR(w) = 1$.
	If $w \not\in \Lang{\PR}$ $\PR(w) = 0$.
	
	Two PRs $\PR_1$ and $\PR_2$ over a common alphabet $\Sigma$ are \emph{equivalent} if $\Lang{\PR_1}$ = $\Lang{\PR_2}$.
	A set (or \emph{language}) $L \subseteq \SP$ is said to be \emph{recognizable} if there exists a PR $\PR$ such that $L = \Lang{\PR}$.
	Intuitively, PRs act as bottom-up deterministic finite tree automata on terms with desirable algebraic properties:
	two terms representing the same pomset share the same acceptance.

	\begin{example}
		\label{ex:PR}
		Let $L$ be the language containing singleton $c$ and every pomset $(a \parallel bu)$ where $u \in L$, i.e. $L = \{c, a \parallel (bc), a \parallel (b (a \parallel (bc))), \dots\}$.
		
		This language is accepted by the PR $\PR = (R, \odot, \obar, \neut, i, F)$ where 
		$R = \{r_a, r_b, r_c, r_{bc}, r_0, \neut\}$,
		$i(x) = r_x$ for $x \in \{a, b, c\}$,
		$F = \{r_c\}$, and
		$\obar$ and $\odot$ are such that $r_b \odot r_c = r_{bc}$, $r_a \obar r_{bc} = r_c$,
		$\neut$ is the neutral element for both operations, and all the other possible products return $r_0$.
	\end{example}

	\subsection{Contexts}
	
	We define pomset patterns potentially featuring placeholder symbols denoted $\square_j$ in lieu of letters that can be replaced by pomsets.
	
	\begin{definition}[Multi-contexts]
		For $m \in \integers \setminus \{0\}$, let $\Xi = \{\square_1, \dots, \square_m\}$ be a set of $m$ distinct letters such that $\Xi \cap \Sigma = \emptyset$.
		The set of \emph{$m$-contexts} $\Ct{m}$ is the subset of $\mathrm{SP}(\Sigma \cup \Xi)$ of pomsets containing exactly one element labelled by~$\square_j$ for all $j \in \{1, \ldots, m\}$.
	\end{definition}
	
	Given $c \in \Ct{m}$ and $w_1, \dots, w_m \in \SP$, $c[w_1, \dots, w_m]$ denotes the pomset where $\square_j$ has been replaced by $w_j$, that is, a term $t_j \in \ST{w_j}$ has been inserted in a term of $t_c \in \ST{c}$ in place of $\square_j$.
	The latter operation trivially yields the same pomset regardless of the representatives $t_j$ and $t_c$ chosen.
	
	We write $\SP = \Ct{0}$.
	We simply call 1-contexts \emph{contexts}, and always denote their placeholder symbol $\square$.
	Given $c_1, c_2 \in \Ct{1}$, $c_1[c_2] \in \Ct{1}$ stands for the context obtained by replacing $\square$ with $c_2$ in $c_1$.
	Context $c_2$ is then said to be a \emph{subcontext} of $c_1[c_2]$.
	
	For $w \in \SP$, a \emph{split} of $w$ is a pair $(c, t) \in \Co \times \ST{z}$ for some $z \in \SP$ such that $w = c[z]$.
	Note that $z$ is a subpomset of $w$.
	Given $C \subseteq \Ct{1}$ and $A \subseteq \SP \cup \Ct{1}$, we define the set $C[A] = \{c[z] \mid c \in C, z \in A\}$. 		
	
%	Similar pomsets w.r.t. a PR remain so after being inserted in a context:
%	\begin{lemma}[{\cite[Lem. 29]{DBLP:conf/fossacs/HeerdtKR021}}]
%		\label{lem:free_context}
%		Given a PR $\PR = (M, \odot, \obar, \neut, i, F)$, for all $w_1, w_2 \in \SP$, if $\eval(w_1) = \eval(w_2)$ then for all $c \in \Ct{1}$, $\eval(c[w_1]) = \eval(c[w_2])$.
%	\end{lemma}

	\subsection{A Myhill-Nerode theorem}
	
	Given $L \subseteq \SP$, we consider the \emph{congruence relation} $\sim_L$ such that 
	for $u, v \in \SP$, $u \sim_L v$ if and only if for all $c \in \Ct{1}$, $c[u] \in L \iff c[v] \in L$.
	It is an equivalence relation on $\SP$ compatible with $\cdot$ and $\parallel$.
	Set $[w]_{\sim_L}$ denotes the equivalence class of $w$ in the quotient space $\mnqspace{L}$.
	There exists a Myhill-Nerode characterization of recognizable languages of $\SP$:
	
	\begin{theorem}[\cite{LW00:sp}]
		\label{th:myhill_nerode}
		A language $L \subseteq \SP$ is recognizable if and only if $\sim_L$ has finite index,  i.e. $\mnqspace{L}$ is finite.
	\end{theorem}
	
	Given a pomset language $L$ and $w_1, w_2 \in \SP$, we say that $c \in \Ct{1}$ is a \emph{distinguishing context} in $L$ for $w_1$ and $w_2$ if $c[w_1] \in L \iff c[w_2] \notin L$, that is, one of $c[w_1]$ and $c[w_2]$ is in $L$ while the other is not.
	If $L$ is recognized by a PR $\PR$, this necessarily implies that $m_1 = \eval(c[w_1]) \neq m_2 = \eval(c[w_2])$: one state must be in $F$ while the other is not.
	We then say that $c$ \emph{distinguishes} states $m_1$ and $m_2$ (resp. pomsets $w_1$ and $w_2$).
	If there is no such $c$, $m_1$ (resp. $w_1$) and $m_2$ (resp. $w_2$) are said to be \emph{indistinguishable}.
	We write ${\sim_\PR} = {\sim_{\Lang{\PR}}}$.
	
	\begin{example}
		\label{ex:context}
		Consider Example~\ref{ex:PR} again.
		Context $\square \parallel c$ distinguishes $a$ from $\varepsilon$, hence states $r_a$ and $e$, as $a \parallel c \notin L$ but $\varepsilon \parallel c = c \in L$.
	\end{example}
		
	\begin{definition}[Reachable, distinguished, minimal pomset recognizers]
		A pomset recognizer $\PR = (R, \odot, \obar, 1, i, F)$ is said to be \emph{reachable} if, for all $m \in R$, there exists $w \in \SP$ such that $\eval(w) = m$; $w$ is said to be an \emph{access pomset} of $m$.
		$\PR$ is \emph{distinguished} if for all $w_1, w_2 \in \SP$ such that $\eval(w_1) \neq \eval(w_2)$, there exists $c \in \Ct{1}$ such that $\PR(c[w_1]) \neq \PR(c[w_2])$.
		$\PR$ is \emph{minimal} if it is both reachable and distinguished.
	\end{definition}

	If $L$ is recognizable, $\sim_L$ induces an obvious, unique minimal pomset recognizer $\PR_L = (\mnqspace{L}, \cdot, \parallel, [\varepsilon]_{\sim_L}, i_L, F_L)$ such that $\forall a \in \Sigma$, $i_L(a) = [a]_{\sim_L}$, $F_L = \{[w]_{\sim_L} \in \mnqspace{L} \mid w \in L\}$ and $[w_1]_{\sim_L} \circ [w_2]_{\sim_L} = [w_1 \circ w_2]_{\sim_L}$ for $\circ \in \operators$.
	The resulting evaluation function $\eval$ is the syntactic homomorphism $\SP \to \mnqspace{L}$ that matches pomsets to their congruence class.
	As a consequence, knowledge of a minimal PR that recognizes $L$ and knowledge of $L$'s congruence relation and classes are one and the same.

	\section{An Active Learning Algorithm for Pomset Recognizers}
	\label{sec:PLlambda}
	
	Consider a recognizable pomset language $L$ on an alphabet $\Sigma$.
	Let the \emph{model} $\Mod$ be the unique minimal PR such that $\Lang{\Mod} = L$.
	\emph{Active learning} (AL) is a game between a learner and a \emph{minimally adequate teacher} (MAT) that consists in the former guessing $\Mod$ by asking two types of queries to the latter:
	\begin{description}
		\item[Membership queries.] Given $w \in \SP$, does $w \in L$, i.e. what is $\Mod(w)$?
		
		\item[Equivalence queries] Given a pomset recognizer $\Hyp$ (called the hypothesis) on $\Sigma$, does $\Lang{\Hyp} = L = \Lang{\Mod}$?
		If it does not, return a \emph{counterexample} $w \in \SP$ such that $\Hyp(w) \neq \Mod(w)$.
	\end{description}
	
	We introduce in this section $\PLlambda$, a new AL algorithm for PRs.

	\subsection{Data structures}
	\label{sec:data}
	
	The ability to infer $\Mod$ from queries stems from Theorem \ref{th:myhill_nerode}.
	AL algorithms compute an \emph{under-approximation} $\sim_\Hyp$ of ${\sim_L} = {\sim_\Mod}$ such that $w_1 \not\sim_\Hyp w_2 \implies w_1 \not\sim_L w_2$.
	Obviously, $w_1 \sim_\Hyp w_2 \implies w_1 \sim_L w_2$ may not hold if the hypothesis is too coarse, thus, $\Hyp$ may have to be refined several times.
	Nevertheless, each refinement increases the number of congruence classes of $\sim_L$ distinguished by $\sim_\Hyp$, until the classes of $\sim_\Hyp$ are exactly the classes of $\sim_L$, at which point ${\sim_\Hyp} = {\sim_\Mod} = {\sim_L}$ and $\Lang{\Hyp} = \Lang{\Mod} = L$;
	$\PLlambda$ then terminates.
	
	\paragraph{Covering the congruence classes.}
	
	$\PLlambda$ maintains a finite set $S$ of pomsets called the set of \emph{access pomsets}, meant to store representatives of $\sim_L$'s congruence classes.	
	By design, $S$ will be closed by the subpomset relation and contain the empty pomset $\varepsilon$.
	The classes of $\sim_L$ are the states of the model $\Mod$.
	
	However, knowledge of its evaluation function $i_\Mod$ and its internal operations $\cdot_\Mod$ and $\parallel_\Mod$ is also required to fully infer $\Mod$.
	Thus, we also introduce a \emph{frontier} set $S^+ = \left(\Sigma \cup \{u \circ v \mid \circ \in \{\cdot,\parallel\}, u, v \in S\}\right) \setminus S$ that contains single letters and combinations of elements of $S$ so that we can infer their congruence classes.
	
	A \emph{pack of components} $\B = \{B_1, \ldots, B_m\}$ partitions $S \cup S^+$ in such a manner that each component contains at least one $s \in S$.
	For $s \in S \cup S^+$, $\B_s$ stands for the only component of $\B$ $s$ belongs to.
	Given $B \in \B$, $\as_\B(B) = S \cap B$ is called the set of \emph{access pomsets} of $B$.
	For $s \in S \cup S^+$, we define $\as_\B(s) = \as_\B(\B_s)$.
	$\B$'s purpose is to under-approximate the classes of $\sim_\Mod$ and classify elements of $S \cup S^+$.
	
	\paragraph{Distinguishing the congruence classes.}
	
	To compute $\B$ and posit which class of $\sim_L$ an element of $S \cup S^+$ belongs to, we introduce a \emph{discrimination tree} (DT) $\D$.
	It is a full binary tree (see Figures~\ref{fig:hyp_1} and \ref{fig:pack_2} for examples):
	its inner nodes are labelled by contexts in $\Co$;
	in particular, its root is labelled by $\square$;
	its leaves are either unlabelled or labelled by a component of $\B$, in such a fashion $\D$'s set of labelled leaves is in bijection with $\B$.
	
	The labels of $\D$'s inner nodes form a set of contexts $\C$.
	Given $B \in \B$, $\C_B$ is defined as the set of contexts that appear along the branch that runs from the root of $\D$ to the leaf labelled by $B$.
	Note that for all $B \in \B$, $\square \in \C_B$.
		
	For any pomset $w \in \SP$, we define the \emph{sifting} operation of $w$ through $\D$:
	starting from the root of $\D$, at every inner node of $\D$ labelled by a context $c$, we perform a MQ on $c[w]$ then branch towards the right (resp. left) child if $c[w] \in L$ (resp. $c[w] \not\in L$).
	We iterate this procedure until a leaf is reached:
	the matching component $B \in \B$ is the result of the sifting operation.
	We define $\D(w) = B$.	
	$\D(w)$ may be undefined if $w$ is sifted into an unlabelled leaf, at which point a new component with access pomset $w$ is created (see Algorithm~\ref{algo:expand}).
	Thus, $\D$ can be viewed as a partial function $\SP \to \B$.
	
	Sifting requires a number of MQs bounded by the height of $\D$.
	Intuitively, DTs are used to classify pomsets:
	pomsets that behave similarly w.r.t. the finite set of distinguishing contexts $\C_B$ are sifted into the same component $B \in \B$;
	thus, $\B_w = \D(w)$.
	Conversely, if two elements of $S \cup S^+$ belong to two different components $B_1$ and $B_2$, then there are distinguished by the deepest common ancestor of leaves $B_1$ and $B_2$ in $\D$.
	
	\begin{example}
		Let us sift $b$ through the DT shown in Figure~\ref{fig:pack_2}.
		First, we query $\square[b] = b$.
		It does not belong to $L$;
		we branch left then query $(\square \cdot c)[b] = bc$.
		The answer is negative and we branch left again.
		Finally, we reach a leaf labelled by $\vert[B_a] = \D(b)$, having performed two MQs.
	\end{example}
	
	\paragraph{Properties of the partition.}
	
	$\B$ is said to be \emph{consistent} if for any $B_1, B_2 \in \B$, $u_1, v_1 \in \as_\B(B_1)$, $u_2, v_2 \in \as_\B(B_2)$, and $\circ \in \operators$, $u_1 \circ u_2$ and $v_1 \circ v_2$ belong to the same component of $\B$:
	no matter the elements of $\as_\B(B_1)$ and $\as_\B(B_2)$ we consider, their composition must belong to the same component.
	Moreover, $\B$ is \emph{$\circ$-associative} for $\circ \in \operators$ if for any $s_1, s_2, s_3 \in S$ and $s_l \in \as_\B(s_1 \circ s_2)$, $s_r \in \as_\B(s_2 \circ s_3)$, $\B_{s_l \circ s_3} = \B_{s_1 \circ s_r}$.
	The \emph{sharpness index} of $\B$ is the smallest integer $\Sindex(\B)$ such that for any $B \in \B$, $|S \cap B| \leq \Sindex(\B)$.
	$\B$ is said to be \emph{sharp} if $\Sindex(\B) = 1$.
	
	If $\B$ is consistent and associative, we can extend the operators $\cdot$ and $\parallel$ to components of $\B$, and the resulting laws will be internal and associative.
	Finally, for any $B \in \B$, since $\square \in \C_B$ and for any $u \in B$, $\square[u] = u$, $\Mod$ is constant on $B$:
	this shared value is written $\Mod(B)$.

	\subsection{Building the hypothesis}
	\label{sec:build_hyp}
	
	\paragraph{Defining the hypothesis.}
	
	If $\B$ is \emph{consistent}, \emph{$\cdot$-associative}, and \emph{$\parallel$-associative}, then we design a minimal hypothesis $\Hyp = (H, \cdot_\Hyp, \parallel_\Hyp, \neut_\Hyp, i_\Hyp, F_\Hyp)$ as follows:
	\begin{itemize}
		\item $H = \B$. $\Hyp$'s states are the postulated congruence classes of $\sim_L$.
		
		\item Given $u, v \in S$, since $\B$ is \emph{consistent}, we can define $\B_u \cdot_\Hyp \B_v = \B_{u \cdot v}$ (resp. $\B_u \parallel_\Hyp \B_v = \B_{u \parallel v}$). We use $S^+$ and $\B$ to build $\Hyp$'s internal operations.
		
		\item $\neut_\Hyp = \B_\varepsilon$. The neutral element is the class of the empty pomset.
		
		\item Given $a \in \Sigma$, $i_\Hyp(a) = \B_a$. We rely on $\Sigma \subseteq S \cup S^+$ to build a pomset homomorphism.
		
		\item $F_\Hyp = \{B \in \B \mid \Mod(B) = 1\}$. A component is accepting if its members are accepted by $\Mod$. $F_\Hyp$ contains the labelled leaves of $\D$ belonging to its right subtree.
	\end{itemize}
	
	For any $w \in \SP$, the state $\eval_\Hyp(w)$ pomset $w$ evaluates to is a component of $\B$ we denote $\B_w$;
	its set of access pomsets is $\as_\Hyp(w) = \as_\B(\B_w)$.
	This notation is compatible with the earlier definition of $\B_w$ for $w \in S \cup S^+$ due to $\Hyp$ being reachable.
	The hypothesis handles pomsets and their access pomsets similarly:
	\begin{property}[Substitution by access pomsets]
		\label{prop:freeness_as}
		$\forall c \in \Co$, $\forall w \in \SP$, $\forall p \in \as_\Hyp(w)$, $\Hyp(c[w]) = \Hyp(c[p])$ and $\eval_\Hyp(c[w]) = \eval_\Hyp(c[p])$.
	\end{property}

	\begin{example}
		\label{ex:hyp_1}
%		Figure~\ref{fig:hyp_1} shows on the left a consistent and associative pack of components $\B = \{\rose[B_\varepsilon], \bleu[B_c]\}$, and on the right the resulting hypothesis $\Hyp_1$.
%		The initialization of $S$, $S^+$, $\D$, and $\B$ by Algorithms~ \ref{algo:expand} and \ref{algo:main} will be discussed later in Section~\ref{sec:algos}.
		Let us learn the pomset recognizer of Example~\ref{ex:PR}.
		As discussed later in Section~\ref{sec:algos}, initially, $S = \{\varepsilon\}$ and the DT $\D$ features a single inner node labelled by $\square$.
		As $\varepsilon \not\in L$, $\D$'s left child is labelled by $\rose[B_\varepsilon]$.
		Sifting the elements of $S^+$ through $\D$ results in a new component $\bleu[B_c]$ being created and $S$, $S^+$ being updated accordingly.
		The resulting $\B$ is consistent and associative, yielding a first hypothesis $\Hyp_1$ (see Figure~\ref{fig:hyp_1}).
		\begin{figure}[H]
			\centering
			\begin{minipage}{0.29\linewidth}
				$S = \{\varepsilon, c\}$ \\
				$S^+ \setminus S = \{a, b, cc, c \parallel c\}$ \\
				$\rose[B_\varepsilon] = \{\boxed{\varepsilon}, a, b, cc, c \parallel c\}$ \\
				$\bleu[B_c] = \{\boxed{c}\}$
			\end{minipage}
			\hfill
			\begin{minipage}{0.18\linewidth}
				\centering
				\begin{tikzpicture}[sibling distance=15mm, level distance=1cm]
					\node {$\square$}
					child {node {$\rose[B_\varepsilon]$}}
					child {node {$\bleu[B_c]$}};
				\end{tikzpicture}
			\end{minipage}
			\hfill
			\begin{minipage}{0.49\linewidth}
				$H = \{\rose[B_\varepsilon], \bleu[B_c]\}$, $F_\Hyp = \{\bleu[B_c]\}$, \\
				$i_\Hyp(a) = i_\Hyp(b) = \rose[B_\varepsilon]$, $i_\Hyp(c) = \bleu[B_c]$, \\
				$\rose[B_\varepsilon] \circ \bleu[B_c] = \bleu[B_c] \circ \rose[B_\varepsilon] = \bleu[B_c]$ and \\ $\rose[B_\varepsilon] \circ \rose[B_\varepsilon] = \bleu[B_c] \circ \bleu[B_c] = \rose[B_\varepsilon]$ for $\circ \in \{\cdot_\Hyp, \parallel_\Hyp\}$.
			\end{minipage}
			\caption{First pack of components, a discrimination tree, and the resulting hypothesis.}
			\label{fig:hyp_1}
		\end{figure}
	\end{example}
	
	\paragraph{Compatibility of the hypothesis.}
	
	Given a set $X \subseteq \SP$ of pomsets, hypothesis $\Hyp$ is $X$-\emph{compatible} if for any $w \in X$, $\Hyp(w) = \Mod(w)$.
	$\Hyp$ is said to be \emph{compatible} with $\B$ if it is compatible with $\underset{B \in \B}{\bigcup} \{c[s] \mid s \in B, c \in \C_B\}$.
	
	AL algorithms such as TTT~\cite{isberner14ttt}, $L^\#$~\cite{vaandrager22lsharp}, or van Heerdt et al.'s adaptation of $L^*$~\cite{DBLP:conf/fossacs/HeerdtKR021} to PRs may not always immediately result in a compatible hypothesis.
	However, incompatibilities provide a `free' counterexample $c[s]$ such that $\Hyp(c[s]) \neq \Mod(c[s])$ without requiring an extra MQ or EQ due to $\Mod(c[s])$ having already been queried when $s$ was sifted into $\B_s$.
	
	Note that compatibility of $\Hyp$ implies its minimality, as shown in Appendix~\ref{app:properties}.

	\subsection{Adapting the $\Llambda$ Algorithm}
	\label{sec:algos}
	
	Inspired by $\Llambda$~\cite{DBLP:conf/birthday/HowarS22}, we outline in this section algorithm $\PLlambda$ and detail in Figure~\ref{fig:algos} the various functions that compose it.
	
	\paragraph{Updating data structures.}
	
	$S$, $S^+$, $\B$, and $\D$ are updated through calls to two functions:
	\texttt{Expand} and \texttt{Refine}, respectively Algorithms~\ref{algo:expand} and \ref{algo:refine} of Figure~\ref{fig:algos}.
	
	Algorithm~\ref{algo:expand} inserts a new pomset $w$ belonging to $S^+$ or equal to $\varepsilon$ into the set~$S$ of access pomsets then updates $S^+$ and $\B$, generating a new component if a pomset is sifted into an unlabelled leaf of $\D$ which is then updated accordingly.
	
	Algorithm \ref{algo:refine} refines a component $B$ into two new components $B_0$ and $B_1$ (thus increasing $|\B|$), assuming a context~$c$ distinguishes two access pomsets of~$B$.
	$S$, $\B$ and $\D$ are updated accordingly.
	In particular, $\D$'s leaf labelled by $B$ is replaced by an inner node labelled by $c$ with children $B_0$ and $B_1$.
	
	Intuitively, Algorithm~\ref{algo:expand} is used to expand the set of access pomsets (hence, the frontier as well) while Algorithm~\ref{algo:refine} creates a new congruence classes by using a context to refine an existing class.
	
	\paragraph{Finding conflicts.}
	
	Consistency and associativity issues result in a new distinguishing context being inductively built from an existing element of $\C$ by functions \texttt{MakeConsistent} (Algorithm~\ref{algo:make_consistent}) or \texttt{MakeAssoc} (Algorithm~\ref{algo:make_assoc}).
	
	Algorithm \ref{algo:make_consistent} refines partition $\B$ whenever it encounters a consistency issue, e.g. class $B$ contains two representatives $p_1$ and $p_2$ such that $p_1 \circ p$ and $p_2 \circ p$ in $S \cup S^+$ for some $p \in S$ and $\circ \in \operators$ do not belong to the same component.
	This inconsistency yields a context $c[\square \circ p]$ that distinguishes $p_1$ and $p_2$, where $c \in \C$ is the label of the deepest common ancestor in $\D$ of $p_1 \circ p$ and $p_2 \circ p$.
	This algorithm returns Boolean $\top$ if and only if $\B$ was already consistent in the first place.
	For brevity, we omit a similar symmetric case involving $p \circ p_1$ and $p \circ p_2$.
	
	Similarly, Algorithm \ref{algo:make_assoc} refines partition $\B$ whenever it encounters an associativity issue and returns Boolean $\top$ if and only if $\B$ is already associative.
	
	\paragraph{Lazy refinement.}
	
	If no consistency or associativity issue is found, $\PLlambda$ builds hypothesis $\Hyp$ defined in Section~\ref{sec:build_hyp} then submits an EQ to the teacher.
	The answer is either positive and the algorithm ends with $\Hyp = \Mod$, or negative and a counterexample $w$ is returned;
	a new algorithm \FindEBP detailed in Section~\ref{sec:find_ebp} infers from $w$ a $c \in \Co$ and $p \in S^+$ such that $c$ distinguishes $p$ from a $p' \in \as_\B(p)$, hence proving that class $B_p$ should be refined.
	
	However, rather than immediately using $c$ to refine $B_p$, $\PLlambda$ performs a signature \emph{delayed refinement}:
	$p$ is added to $S$, $S^+$ and $\B$ are updated accordingly, but $c$, being a context of arbitrary size w.r.t. $\Mod$, is not inserted in $\D$.
	This optimization leads to significant improvements of the algorithm's symbol complexity that we discuss in Sections~\ref{sec:complexity} and \ref{sec:exp}.
	
	Instead, function \texttt{AnalyzeCE} (Algorithm~\ref{algo:analyze_ce}) 
	waits for an associativity or consistency defect to necessarily arise in order to inductively update $\C$ and $\D$, rather than directly use $c$.
	Moreover, $c[p]$ and $c[p']$ are added to a counterexample pool $\ce$ (Line~\ref{algo:analyze_ce:refine_later}) for any representative $p' \in \as_\Hyp(p)$;
	indeed, as long as $p$ and $p'$ are not distinguished, either $c[p]$ or $c[p']$
	will remain a counterexample:
	if $p, p' \in S$ belong to the same component $B$ then $\Hyp(c[p]) = \Hyp(c[p'])$;
	but $\Mod(c[p]) \neq \Mod(c[p'])$.
	This loop eventually depletes $\ce$ and ends with $\B$ being sharp.

	\newpage
	\newgeometry{margin=3.5cm}
	\begin{landscape}
		\begin{figure}
			\caption{The various components of the $\PLlambda$ algorithm.}
			\label{fig:algos}
				\begin{multicols*}{2}
				\centering
				\scalebox{0.83}{\begin{minipage}{\linewidth}
					\begin{algorithm}[H]
						\caption{$\mathtt{Expand}(w)$ where $w \in S^+$ or $w = \varepsilon$ if $\B = \emptyset$}
						\label{algo:expand}
						\begin{algorithmic}[1]
							\State $S \gets S \cup \{w\}$
							\For{$p \in \{w\} \cup \{p' \circ w, w \circ p' \mid \circ \in \operators, p' \in S\} \cup \Sigma$} \label{algo:expand:letters}
								\If{$p$ does not belong to any class of $\B$}
									\State $B \gets \D(p)$
									\If{$B$ is defined}
										\State $B \gets B \cup \{p\}$
									\Else \label{algo:expand:init_start}
										\State $B_p \gets \{p\}$
										\State $\B \gets \B \cup \{B_p\}$
										\State $\mathtt{UpdateTreeLeaf}(\D, p, B_p)$
										\State $\mathtt{Expand}(p)$ \label{algo:expand:init_end}
									\EndIf
								\EndIf
							\EndFor
						\end{algorithmic}
					\end{algorithm}
				\end{minipage}}
				\vfill
				\scalebox{0.83}{\begin{minipage}{\linewidth}
					\begin{algorithm}[H]
						\caption{$\mathtt{Refine}(B, c)$ where $B \in \B$, $c \in \Co$, and $\exists z_1, z_2 \in B$, $\Mod(c[z_1]) \neq \Mod(c[z_2])$}
						\label{algo:refine}
						\begin{algorithmic}[1]
							\State $B_0 \gets \{w \in B \mid \Mod(c[w]) = 0\}$
							\State $B_1 \gets \{w \in B \mid \Mod(c[w]) = 1\}$
							\State $\B \gets (\B \setminus \{B\}) \cup \{B_0, B_1\}$
							\State $\mathtt{RefineTree}(\D, B, c, B_0, B_1)$ \label{algo:refine:tree}
							\If{$S \cap B_0 = \emptyset$}
								$\mathtt{Expand}(p_0)$ for some $p_0 \in B_0$ \label{algo:refine:b0_no_as}
							\EndIf
							\If{$S \cap B_1 = \emptyset$}
								$\mathtt{Expand}(p_1)$ for some $p_1 \in B_1$ \label{algo:refine:b1_no_as}
							\EndIf
						\end{algorithmic}	
					\end{algorithm}
				\end{minipage}}
				\vfill
				\scalebox{0.83}{\begin{minipage}{\linewidth}
					\begin{algorithm}[H]
						\caption{$\mathtt{MakeConsistent}()$}
						\label{algo:make_consistent}
						\begin{algorithmic}[1]
							\State $\mathtt{already\_consistent} \gets \top$
							\While{$\exists \circ \in \operators$, $\exists B \in \B$, $\exists p_1, p_2 \in \as_\B(B)$, $\exists p \in S$, $\D(p_1 \circ p) \neq \D(p_2 \circ p)$}
								\State Let $c \in \C$ be such that $\Mod(c[p_1 \circ p]) \neq \Mod(c[p_2 \circ p])$.
								\State $\mathtt{Refine}(B, c[\square \circ p])$ \label{algo:make_consistent:refine}
								\State $\mathtt{already\_consistent} \gets \bot$
							\EndWhile
							\State \Return $\mathtt{already\_consistent}$
						\end{algorithmic}	
					\end{algorithm}
				\end{minipage}}
				\newcolumn
				
				\scalebox{0.83}{\begin{minipage}{\linewidth}
					\begin{algorithm}[H]
						\caption{$\mathtt{MakeAssoc}()$}
						\label{algo:make_assoc}
						\begin{algorithmic}[1]
							\State $\mathtt{already\_assoc} \gets \top$
							\While{$\exists \circ \in \operators$, $\exists s_1, s_2, s_3 \in S$, $\exists s_l \in \as_\B(s_1 \circ s_2)$, $\exists s_r \in \as_\B(s_2 \circ s_3$), $\D(s_1 \circ s_r) \neq \D(s_l \circ s_3)$}
								\State Let $c \in \C$ be such that $\Mod(c[s_1 \circ s_r]) \neq \Mod(c[s_l \circ s_3])$.
								\State $\mathtt{query} \gets \Mod(c[s_1 \circ s_2 \circ s_3])$
								\If{$\Mod(c[s_l \circ s_3]) \neq \mathtt{query}$}
									\State \texttt{Refine}$(\B_{s_1 \circ s_2}, c[\square \circ s_3])$ \label{algo:make_assoc:refine_1}
								\Else
									\State \texttt{Refine}$(\B_{s_2 \circ s_3}, c[s_1 \circ \square])$ \label{algo:make_assoc:refine_2}
								\EndIf
								\State $\mathtt{already\_assoc} \gets \bot$
							\EndWhile
							\State \Return $\mathtt{already\_assoc}$
						\end{algorithmic}
					\end{algorithm}
				\end{minipage}}
				\vfill
				\scalebox{0.83}{\begin{minipage}{\linewidth}
					\begin{algorithm}[H]
						\caption{$\mathtt{AnalyzeCE}(w)$ where $w \in \SPplus$ is such that $\Hyp(w) \neq \Mod(w)$}
						\label{algo:analyze_ce}
						\begin{algorithmic}[1]
							\State $\ce \gets \{w\}$
							\While{$\exists u \in \ce, \Mod(u) \neq \Hyp(u)$} \label{algo:analyze_ce:while_cond}
								\State $(c, p) \gets \mathtt{FindEBP}(\square, t)$ where $t$ is a minimal term of $u$
								\State $\ce \gets \ce \cup \{c[p]\} \cup \{c[p'] \mid p' \in \as_\Hyp(p)\}$ \label{algo:analyze_ce:refine_later}
								\State $\mathtt{Expand}(p)$ \label{algo:analyze_ce:new_component}
								\Repeat \label{algo:analyze_ce:consistency_loop_start}
								\Until{$\mathtt{MakeConsistent}() \land \mathtt{MakeAssoc}()$} \label{algo:analyze_ce:consistency_loop_end}
								\State $\Hyp \gets \mathtt{BuildHypothesis}(S, \B)$	\label{algo:analyze_ce:build_hyp}	
							\EndWhile
						\end{algorithmic}
					\end{algorithm}
				\end{minipage}}
				\vfill
				\scalebox{0.83}{\begin{minipage}{\linewidth}
					\begin{algorithm}[H]
						\caption{$\mathtt{Learn}()$}
						\label{algo:main}
						\begin{algorithmic}[1]
							\State $S, \B, \D \gets \emptyset, \emptyset, \mathtt{Tree}(\square)$
							\State $\mathtt{Expand}(\varepsilon)$			
							\State $\Hyp \gets \mathtt{BuildHypothesis}(S, \B)$\label{algo:main:first_hyp}
							\While{$\exists w \in \SP$, $\Hyp(w) \neq \Mod(w)$} \label{algo:main:find_ce}
								\State $\mathtt{AnalyzeCE}(w)$ \label{algo:main:handle_ce}
								\While{$\exists B \in \B$, $\exists s \in B$, $\exists c \in \C_B$, $\Hyp(c[s]) \neq \Mod(c[s])$} \label{algo:main:fix_comp_start}
									\State $\mathtt{AnalyzeCE}(c[s])$ \label{algo:main:fix_comp_end}
								\EndWhile
							\EndWhile
							\State \Return \Hyp
						\end{algorithmic}
					\end{algorithm}
				\end{minipage}}
				\vfill
			\end{multicols*}
		\end{figure}
	\end{landscape}
	\restoregeometry
	
	\paragraph{The main loop.}
	
	Function \texttt{Learn} (Algorithm~\ref{algo:main}) first initializes $S = \{\varepsilon\}$ and $\C = \{\square\}$, then build the pack of components $\B$ and a first hypothesis $\Hyp$ by expanding the empty pomset $\varepsilon$.
	It then submits $\Hyp$ to the teacher.
	
	If the EQ returns a counterexample $w$, a call to \texttt{AnalyzeCE} identifies new components and $\Hyp$ is refined accordingly.
	Otherwise, a model $\Hyp$ equivalent to $\Mod$ has been learnt and the algorithm returns $\Hyp$.
	Lines \ref{algo:main:fix_comp_start} and \ref{algo:main:fix_comp_end} guarantee that $\Hyp$ is compatible before submitting an EQ.
	
	\begin{example}
		\label{ex:refine}
		Consider counterexample $w = (cc)(a \parallel c)$ to Example~\ref{ex:hyp_1} in the context of Example~\ref{ex:PR}'s learning.
		A call to \FindEBP (later described in Example~\ref{ex:find_ebp}) returns context $\square \parallel c$ and pomset $a \in S^+$.
		A subsequent call to \texttt{Expand} results in $a$ being added to $S$ and $S^+$ being consequently extended.
		
		The resulting (second) pack of components (shown in Figure~\ref{fig:pack_2}) is neither sharp nor consistent:
		$\varepsilon, a \in \rose[B_\varepsilon] \cap S$, $\varepsilon \cdot c = c \in \bleu[B_c]$ yet $ac \in \rose[B_\varepsilon]$.
		$\rose[B_\varepsilon]$ is thus refined by $\square[\square \cdot c] = \square \cdot c$, leading to the creation of a new component $\vert[B_a]$.
		
		The resulting (third) pack of component $\B$ is then consistent and associative.
		\begin{figure}[H]
			\centering
			\begin{minipage}{0.37\linewidth}
				$S = \{\varepsilon, c, a\}$ \\ 
				$S^+ \setminus S = \{b, cc, aa, ca, ac, \\ a \parallel a, a \parallel c, c \parallel c\}$ \\
				$\rose[B_\varepsilon] = \{\boxed{\varepsilon}, \boxed{a}, b, cc, aa, ca, ac, \\ a \parallel c, c \parallel c, a \parallel a\}$ \\
				$\bleu[B_c] = \{\boxed{c}\}$
			\end{minipage}
			\hfill
			\begin{minipage}{0.37\linewidth}
				$S = \{\varepsilon, c, a\}$ \\ 
				$S^+ \setminus S = \{b, cc, aa, ca, ac, \\ a \parallel c, a \parallel a, c \parallel c\}$ \\
				$\rose[B_\varepsilon] = \{\boxed{\varepsilon}\}$ \\		
				$\vert[B_a] = \{\boxed{a}, b, cc, aa, a \parallel a, ca, \\ ac, c \parallel c, a \parallel c\}$ \\
				$\bleu[B_c] = \{\boxed{c}\}$
			\end{minipage}
			\hfill
			\begin{minipage}{0.22\linewidth}
				\centering
				\begin{tikzpicture}[sibling distance=15mm, level distance=1cm]
					\node {$\square$}
					child {node {$\square \cdot c$} [sibling distance=10mm]
						child {node {$\vert[B_a]$}}
						child {node {$\rose[B_\varepsilon]$}}}
					child {node {$\bleu[B_c]$}};
				\end{tikzpicture}
			\end{minipage}
			\caption{Second and third packs of components, second discrimination tree.}
			\label{fig:pack_2}
		\end{figure}
	\end{example}

	\section{A New Counterexample Analysis Algorithm}
	\label{sec:find_ebp}
	
	AL algorithms infer from a counterexample $w$ to hypothesis~$\Hyp$ a \emph{refining pair} $(c, p) \in \Co \times S^+$ such that $\exists p' \in \as_\Hyp(p)$, $\Mod(c[p']) \neq \Mod(c[p])$.
	It proves that $p$ and $p'$ are distinguished by $c$ in $\Mod$, yet belong to the same component $\B_p$, thus highlighting a necessary refinement of~$\B_p$.
	Inspired by Drewes et al.~\cite{DBLP:conf/dlt/DrewesH03}, van Heerdt et al.~\cite{DBLP:conf/fossacs/HeerdtKR021} designed \texttt{HandleCounterExample} (from now on, \HCE), an algorithm that infers a refining pair from $w$ in $\bigo(\Sindex(\B) \cdot |w|)$ MQs.

	We introduce in this section one of our main contributions, \FindEBP: a new counterexample analysis algorithm that requires only $\bigo(\Sindex(\B) \cdot \depth{w})$ MQs.

	\subsection{Breaking points}
	
	Inspired by the work of Rivest and Schapire~\cite{DBLP:journals/iandc/RivestS93}, we are trying to infer \emph{breaking points} (BPs) from counterexamples: witnesses to the model and hypothesis no longer being in agreement over a computation.
	Given a consistent, associative $\B$ from which $\Hyp$ is inferred, we extend these definitions to pomsets and terms.
	
	\begin{definition}[Agreement]
		\label{def:agree}
		Given $c \in \Co$ and $z \in \SP$, we define the \emph{agreement} predicate $\agree(c, z) = ``\forall p \in \as_\Hyp(z), \Hyp(c[p]) = \Mod(c[p])"$.
	\end{definition}

	\begin{definition}[Breaking point]
		\label{def:bp}
		Given $w \in \SPplus$, a split $(c, t) \in \Co \times \ST{z}$ of $w$ is an \emph{Effective Breaking Point} (EBP) if $\agree(c, z) = 1$ and either:
		\begin{enumerate}
			\item $z \in S^+$ and $\Hyp(w) \neq \Mod(w)$.
			
			\item $t = t_1 \circ t_2$ for some $\circ \in \operators$, $t_i \in \ST{z_i}$ for $i \in \{1, 2\}$, and:
			\begin{enumerate}
				\item $\agree(c[\square \circ z_2], z_1) = 0$; this equality implies that $\exists p_1 \in \as_\Hyp(z_1)$, $\Hyp(c[p_1 \circ z_2]) \neq \Mod(c[p_1 \circ z_2])$;
				
				\item $\agree(c[p_1 \circ \square], z_2) = 0$.
			\end{enumerate}
			If only $(a)$ holds, the BP is said to be \emph{tentative} (TBP).
		\end{enumerate}
	\end{definition}
	
	Note that computing $\agree(c, z)$ requires at most $\Sindex(\B)$ MQs.
	We seek EBPs to induce a refinement of $\B$:
	
	\begin{lemma}
		If $(c, t)$ is an EBP of $\Hyp$ w.r.t. $\Mod$, then $\B$ admits a refining pair.
	\end{lemma}

	\begin{proof}
		Indeed, $\agree(c, z) = 1$ implies  $\Hyp(c[p']) = \Mod(c[p'])$ for every $p' \in \as_\Hyp(z)$.
		Case 1. (resp. 2.) of Definition~\ref{def:bp} implies $\Hyp(c[p]) \neq \Mod(c[p])$ for $p = z$ (resp. $p = p_1 \circ p_2 \in S^+$ where $\agree(c[p_1 \circ \square], z_2) = 0$ yields a $p_2 \in \as_\Hyp(z_2)$).
		By Property~\ref{prop:freeness_as}, $\Hyp(c[p]) = \Hyp(c[p'])$. 
		Thus, $\Mod(c[p]) \neq \Mod(c[p'])$ and $c$ distinguishes $p$ from $p'$. \hfill\qed
	\end{proof}
	
	\subsection{The algorithm \FindEBP}
	
	Instead of performing a prefix traversal of a term $t$ like \HCE, Algorithm~\ref{algo:find_ebp} recursively descends along a branch of $t$ until
	it finds an EBP, using TBPs to orient itself and prune irrelevant branches.
	
	Indeed, if $z \in S^+$, by Definition~\ref{def:bp}, $(c, t)$ is an EBP.
	In particular, note that $\Sigma \subseteq S \cup S^+$.
	Thus, the algorithm terminates if it reaches a leaf.
	However, if $t = t_1 \circ t_2$ for some $\circ \in \operators$, we first try to determine whether $(c, t)$ is a TBP:
	\begin{itemize}
		\item If $\agree(c[\square \circ z_2], t_1) = 1$, we can recursively call $\mathtt{FindEBP}(c[\square \circ z_2], t_1)$ as its preconditions still hold:
		$t$'s leftmost branch is explored until a TBP is found.
		
		\item However, if $\agree(c[\square \circ z_2], t_1) = 0$, by Definition~\ref{def:agree}, $\exists p_1 \in \as_\Hyp(z_1)$,  $\Hyp(c[p_1 \circ z_2]) \neq \Mod(c[p_1 \circ z_2])$ \textbf{(\romannumeral 1)}.
		We then determine whether $(c, t)$ is effective or not:
		\begin{itemize}
			\item If $\agree(c[p_1 \circ \square], t_2) = 0$, $(c, t)$ is effective and the algorithm ends. % Détails.
			
			\item Otherwise $\agree(c[p_1 \circ \square], t_2) = 1$ and by \textbf{\romannumeral 1}, $c[p_1 \circ z_2]$ is a counterexample.
			We can recursively call $\mathtt{FindEBP}(c[p_1 \circ \square], t_2)$ as its preconditions hold.
			
			Practically speaking, if the BP turns out not to be effective, we prune $t$'s left subterm $t_1$, replace it with an access pomset $p_1$, then switch branch and recursively explore $t$'s right subterm $t_2$ instead.
		\end{itemize}
	\end{itemize}
	
	{\begin{algorithm}
		\caption{$\mathtt{FindEBP}(c, t)$ where $c \in \Co$, $t$ is a minimal term of $z \in \SPplus$, $\Hyp(c[z]) \neq \Mod(c[z])$ and $\agree(c, z) = 1$}
		\label{algo:find_ebp}
		\begin{algorithmic}[1]
			\If{$z \in S^+$} 
				\Return $(c, z)$ \label{algo:find_ebp:base}
				\ElsIf{$t = t_1 \circ t_2$, $t_i$ being a minimal term of $z_i \in \SPplus$ for $i \in \{1, 2\}$}
					\If{$\agree(c[\square \circ z_2], t_1)$} \label{algo:find_ebp:find_bp_start}
						\State \Return $\mathtt{FindEBP}(c[\square \circ z_2], t_1)$ \label{algo:find_ebp:find_bp_end}
					\Else				
						\State Let $p_1 \in \as_\Hyp(z_1)$ be such that $\Hyp(c[p_1 \circ z_2]) \neq \Mod(c[p_1 \circ z_2])$. \label{algo:find_ebp:check_ebp_start}
						\If{$\agree(c[p_1 \circ \square], t_2)$} \label{algo:find_ebp:check_ebp_end}
							\State \Return $\mathtt{FindEBP}(c[p_1 \circ \square], t_2)$ \label{algo:find_ebp:look_for_ebp}
						\Else
							\State Let $p_2 \in \as_\Hyp(z_2)$ be such that $\Hyp(c[p_1 \circ p_2]) \neq \Mod(c[p_1 \circ p_2])$. \label{algo:find_ebp:ebp_found_start}	
							\State \Return $(c, p_1 \circ p_2)$ \label{algo:find_ebp:ebp_found_end}
					\EndIf
				\EndIf
			\EndIf
		\end{algorithmic}
	\end{algorithm}}

	\begin{example}
		\label{ex:find_ebp}
		Let us resume Example~\ref{ex:refine}.
		Consider a call to $\mathtt{FindEBP}(\square, t)$  where term $t \in \ST{cc(a \parallel c)}$ is displayed in Figure~\ref{fig:findebp_1}.
		Initially, we try replacing the left subterm $cc$ rooted in node \textbf{1} with its access pomset $\varepsilon$.
		Since $\agree(\square \cdot (a \parallel c), cc) = 0$, we found a TBP.
		We now try to replace subterm $a \parallel c$ rooted in node \textbf{2} with its access pomset $c$.
		But $\agree(\varepsilon \cdot \square, a \parallel c) = 1$ and the BP is not effective.
		
		We therefore replace $cc$ with $\varepsilon$ and we switch to a new branch, exploring node \textbf{2} by calling $\FindEBP(\varepsilon \cdot \square, a \parallel c)$.
		We try replacing subterm $a$ rooted in node \textbf{3} with its access pomset $\varepsilon$.
		As $\agree(\varepsilon \cdot (\square \parallel c), a) = 1$, we recursively call $\FindEBP(\varepsilon \cdot (\square \parallel c), a)$ and explore node $3$.
		
		We have reached a base case:
		$a \in S^+$ is distinguished from its access pomset $\varepsilon \in S$ by context $\varepsilon \cdot (\square \parallel c) = \square \parallel c$.
		
		\begin{figure}[H]
			\begin{minipage}{0.31\linewidth}
				\centering
				\begin{tikzpicture}[sibling distance=15mm, level distance=0.7cm]
					\node {$\cdot^0$}
					child {node {$\cdot^1$} [sibling distance=11mm]
						child {node {$c$}}
						child {node {$c$}}}
					child {node {$\parallel^2$} [sibling distance=11mm]
						child {node {$a^3$}}
						child {node {$c$}}};
				\end{tikzpicture}
			\end{minipage}
			\hfill
			\begin{minipage}{0.31\linewidth}
				\centering
				\begin{tikzpicture}[sibling distance=15mm, level distance=0.7cm]
					\node[purple] {$\cdot^0$}
					child[gray] {node {$\varepsilon^1$} edge from parent[gray]}
					child[purple] {node {$\parallel^2$} [sibling distance=11mm]
						child[black] {node {$a^3$}}
						child[black] {node {$c$}}};
				\end{tikzpicture}
			\end{minipage}
			\hfill
			\begin{minipage}{0.31\linewidth}
				\centering
				\begin{tikzpicture}[sibling distance=15mm, level distance=0.7cm]
					\node[purple] {$\cdot^0$}
					child[gray] {node {$\varepsilon^1$} edge from parent[gray]}
					child[purple] {node {$\parallel^2$} [sibling distance=11mm]
						child[purple] {node {$a^3$} edge from parent[purple]}
						child[black] {node {$c$}}};
				\end{tikzpicture}
			\end{minipage}
			\caption{Running \FindEBP on counterexample $(cc)(a \parallel c)$ yields $(\square \parallel c, a)$.}
			\label{fig:findebp_1}
		\end{figure}
	\end{example}

	\section{Termination, Correction and Complexity}
	\label{sec:complexity}

	\subsection{Algorithm \FindEBP}
	
	A proof (see Appendix~\ref{app:proof_find_ebp}) of Algorithm~\ref{algo:find_ebp} can be performed by induction on $t$'s depth:
	the base case holds thanks to Line~\ref{algo:find_ebp:base}, while Lines~\ref{algo:find_ebp:find_bp_end} and \ref{algo:find_ebp:look_for_ebp} handle the inductive case, as both calls preserve \FindEBP's preconditions.
	
	If $t$ is of depth $d$, $\mathtt{FindEBP}(\square, t)$ performs at most $d$ recursive calls, and the agreement predicate is called twice per call (Lines~\ref{algo:find_ebp:find_bp_start} and \ref{algo:find_ebp:check_ebp_end}).
	Given a minimal term~$t$ of a counterexample $w$, $\mathtt{FindEBP}(\square, t)$ thus performs $\bigo(\Sindex(\B) \cdot \depth{w})$ MQs.
	Moreover, if $\B$ is sharp, e.g. in the context of van Heerdt et al.'s $\PLstar$, then $\bigo(\depth{w})$ queries are required.
	Finally, if we assume $t$ is a full binary tree, only $\bigo(\log{|w|})$ queries are needed.
	By comparison, \HCE performs $\bigo(\Sindex(\B) \cdot |w|)$ MQs.

	\subsection{The full algorithm $\PLlambda$}
	
	\paragraph{Correction.}
	
	As mentioned in Section~\ref{sec:algos}, correction of $\PLlambda$ stems from $\sim_{\Hyp}$ being an under-approximation of $\sim_{\Mod}$ on $S \cup S^+$:
	the discrimination tree $\D$ explicitly distinguishes the components of $\B$.
	Once both congruence relations have the same index over $S \cup S^+$, we can fully infer $\Mod$ from $\B$.
	
	\paragraph{Termination.}
	
	Algorithms~\ref{algo:make_consistent} and \ref{algo:make_assoc} refine $\B$ each time a consistency or associativity defect is found.
	By design, $\B$ features at most $|\mnqspace{\Mod}|$ classes:
	thus, these algorithms terminate with a consistent, associative pack of components.
	
	In a similar fashion, each call to $\mathtt{FindEBP}(\square, t)$ in Algorithm~\ref{algo:analyze_ce} returns a new refining pair.
	Thus, only a finite number of iterations of the \textbf{while} loop can be performed, and Algorithm~\ref{algo:analyze_ce} terminates.
	Moreover, it ends with $\B$ being sharp:
	if $p, p' \in S$ belong to the same component $B$, $\Hyp(c[p]) = \Hyp(c[p'])$; but $\Mod(c[p]) \neq \Mod(c[p'])$, hence either $c[p]$ or $c[p']$ remains a counterexample.

	Finally, as each counterexample returned by the teacher results in at least one new class of $\sim_L$ being identified, Algorithm~\ref{algo:main} ends.
	
	\paragraph{Query complexity.}
	
	Let $|\Mod| = n$, $\sigma = |\Sigma|$, $m$ and $d$ the maximal size and depth of counterexamples returned by the MAT.
	Let us analyze $\PLlambda$'s complexity.
	
	$\PLlambda$ performs at most $\bigo(n)$ EQs.
	Thus, there are at most $n$ calls to \FindEBP, and each call needs $\bigo(d \cdot n)$ MQs:
    due to $\B$ possibly not being sharp, $1 \leq \Sindex(\B) \leq n$.
	Computing $\B$ requires $\bigo(n \cdot (n^2 + \sigma))$ MQs:
	since $|\C| \leq n$, sifting a pomset requires at most $n$ queries, and $|S| \leq n$, $|S^+| \leq n^2 + \sigma$.
	
	Finally, $\PLlambda$ with \FindEBP performs $\bigo(n \cdot (n^2 + \sigma) + d \cdot n^2)$ MQs, whereas $\PLstar$~\cite{DBLP:conf/fossacs/HeerdtKR021} with \HCE requires $\bigo(n \cdot (n^2 + \sigma) + m \cdot n)$ queries, or $\bigo(n \cdot (n^2 + \sigma) + d \cdot n)$ were we to replace \HCE with \FindEBP.
	Similarly to $\Llambda$~\cite{DBLP:conf/birthday/HowarS22}, we show that this unfavourable comparison does not empirically hold in Section~\ref{sec:exp}.
	
	\paragraph{Symbol complexity.}

	A mere estimate of the number of queries overlooks the fact that the actual execution time of a MQ depends on the size of its input.
	Thus, we consider the \emph{symbol complexity} of $\PLlambda$, i.e. the sum of the sizes of the pomsets queried.
	Since $S$ is built inductively from elements of $S^+$, for any $s \in S \cup S^+$, $|s| = \bigo(2^n)$.
	Similarly, $\C$ is built inductively by combining some existing element of $\C$ with context $\square \circ s$ for some $s \in S$ and $\circ \in \operators$;
	thus, for any $c \in \C$, $|c| = \bigo(2^n \cdot n)$.
	A query $\Mod(c[s])$ is therefore of size $\bigo(\bleu[2^n \cdot n])$.
	Computing $\B$ thus requires $\bigo(\bleu[2^{n} \cdot n^2 \cdot (n^2 + \sigma)])$ symbols.
	
	The pomset inductively manipulated by \FindEBP is of size $\bigo(\rouge[m + d \cdot 2^n])$, and so are queries performed by \FindEBP:
	out of the original term of size $m$, up to one subterm per level may be replaced by an access pomset of size at most $2^n$.
	Assuming it relies on \FindEBP, $\PLlambda$'s symbol complexity is therefore
	$\bigo(\bleu[2^{n} \cdot n^2  \cdot (n^2 + \sigma)] + \rouge[d \cdot n^2 \cdot (m + d \cdot 2^n)])$.
	
	By comparison, $\PLstar$ infers its contexts directly from counterexamples:
	they are therefore of size $\bigo(m)$, and a query $\Mod(c[s])$, of size $\bigo(m + 2^n)$.
	$\PLstar$'s symbol complexity is $\bigo(\bleu[(m + 2^n) \cdot n \cdot (n^2 + \sigma)] + \rouge[d \cdot n \cdot (m + d \cdot 2^n)])$ if we assume it also relies on \FindEBP.
	Thus, the \bleu[symbol complexity of $\PLlambda$'s computation of $\B$] does not depend on $m$, although the \rouge[counterexample analysis procedure] obviously does.
	This property results in sizeable complexity gains, as shown in Section~\ref{sec:exp}.

	\section{Generating Test Suites for Equivalence Queries}
	\label{sec:wmethod}
	
	%	The use of EQs in active learning algorithms is paradoxical:
	%	to infer a formal model from a black box, we must be able to submit EQs to the very model we are trying to learn.
	We remedy the use of EQs by designing a suitable finite test suite ensuring general equivalence between two pomsets recognizers,
	%	Naturally, this is not possible in the general case.
	assuming that the size of the system is bounded w.r.t. the hypothesis we submit.
	This test suite extends the $W$-method~\cite{DBLP:journals/tse/Chow78,Vasilevskii1973,Moerman19} to recognizable languages of pomsets.
	
	\begin{definition}[Equivalence on a test suite]
		Let $Z \subseteq \SP$.
		Two PRs $\PR_1$ and $\PR_2$ are said to be \emph{$Z$-equivalent}, written $\Hyp \equiv_Z \Mod$, if for any $z \in Z$, $\Hyp(z) = \Mod(z)$.
	\end{definition}
	
	Given a hypothesis $\Hyp = (H, \cdot_\Hyp, \parallel_\Hyp, \neut_\Hyp, \eval_\Hyp, F_\Hyp)$ of size  $|H| = n$, and a model $\Mod = (M, \cdot_\Mod, \parallel_\Mod, \neut_\Mod, \eval_\Mod, F_\Mod)$ sharing the same alphabet such that $\Hyp$ and $\Mod$ are minimal, assume that a bound $k$ such that $0 \leq |M|-|H| \leq k$ is known.

	\subsection{Computing a state cover}
	
	We seek a test suite that covers and distinguishes the states of $\Mod$.
	
	\begin{definition}[State cover]
		A set $P \subseteq \SP$ is a \emph{state cover} of a PR $\PR = (R, \odot, \obar, \neut, i, F)$ if $\varepsilon \in P$ and every $r \in R$ admits an access pomset $p \in P$.
	\end{definition}
	
	\begin{definition}[Characterization set]
		A set of contexts $W \subseteq \Ct{1}$ is a \emph{characterization set} of a PR $\PR = (R, \odot, \obar, \neut, i, F)$ if $\square \in W$ and for any $r_1, r_2 \in R$, if $r_1$ and $r_2$ are distinguishable, then $\exists c \in W$, $\PR(c[r_1]) \neq \PR(c[r_2])$.
	\end{definition}
	
	Based on bound $k$ and our knowledge that states distinguished in $\Hyp$ are still distinguished in $\Mod$, we use $P$ to design a state cover of the unknown PR $\Mod$.
	
	\begin{definition}[Extended state cover]
		Given a state cover $P$ of $\Hyp$, it is the set $\Lcov{i}{P} = \{c[p] \mid m \in \integers, c \in \Ct{m}, \depth{c} \leq i, p \in P^m\}$.
	\end{definition}
	
	Intuitively, $\Lcov{i}{P}$ consists of all pomsets obtained by inserting access pomsets of $P$ in a multi-context of height equal to or smaller than $i$. Then:
	\begin{theorem}
		\label{th:canonical_state_cover}
		Let $P$ be a state cover of $\Hyp$ and $W$ a characterization set of $\Hyp$ such that $\Hyp \equiv_{W[P]} \Mod$. Then $\Lcov{k}{P}$ is a state cover of $\Mod$.		
	\end{theorem}
	
	The proof of this theorem is detailed in Appendix~\ref{app:proof_state_cover}.
	Intuitively, branches of an access term to a state of $\Mod$ can be shortened in such fashion subterms below depth $k$ are replaced by access terms to $\Hyp$.
	%	First, since $\Hyp$ and $\Mod$ are $W[P]$-equivalent, $P$ covers a set $M_P$ of at least $n$ demonstrably distinguishable states of $\Mod$.
	%	Then, given any representative $w$ of a state of $\Mod$, by removing repeating patterns from $w$'s evaluation tree and replacing subterms that reach states in $M_P$ by their matching access pomset in $P$, we can design a representative of the same state that belongs to $\Lcov{k}{P}$.

	\subsection{Exhaustivity of the test suite}
	
	Using $P$ and $W$, our goal is to design a \emph{complete} test suite $Z$, i.e. such that $Z$-equivalence must imply full equivalence of $\Hyp$ and $\Mod$.
	To do so, we extend the notion of bisimulation to PRs:
	\begin{definition}[Bisimulation relation]
		A \emph{bisimulation relation} $\bisim$ between two PRs $\PR_1 = (R_1, \odot_1, \obar_1, \neut_1, i_1, F_1)$ and $\PR_2 = (R_2, \odot_2, \obar_2, \neut_2, i_2, F_2)$ is a binary relation $R_1 \times R_2$ such that:
		\begin{enumerate}
			\item $r_1 \bisim r_2$ implies that $r_1 \in F_1 \iff r_2 \in F_2$.
			\item $r_1 \bisim r_2$ and $r'_1 \bisim r'_2$ implies that $r_1 \circ_1 r_2 \bisim r'_1 \circ_2 r'_2$ for $\circ \in \{\odot, \obar\}$.
		\end{enumerate}
	\end{definition}
	
	% Similarly to finite automata, bisimilarity implies language equivalence:
	
	\begin{lemma}
		\label{prop:bisim_equiv}
		Given a bisimulation relation $\bisim$ between $\PR_1$ and $\PR_2$, if $\PR_1$ and $\PR_2$ share the same alphabet $\Sigma$, the same neutral element $\neut$, and for any $x \in \Sigma \cup \{e\}$, $\eval_{\PR_1}(x) \bisim \eval_{\PR_2}(x)$, then $\PR_1$ and $\PR_2$ are equivalent.
	\end{lemma}
	
	We design a set $Z$ such that that $Z$-equivalence results in bisimilarity:
	\begin{theorem}
		Let $P$ be a minimal (w.r.t. $\subseteq$) state cover of $\Hyp$, $W$ be a minimal  characterization set of $\Hyp$ and $Z = W[\Lcov{k+1}{P}]$.
		We introduce the binary relation $\bisim$ on $H \times M$:
		$${\bisim} = \{(\eval_\Hyp(l), \eval_\Mod(l)) \mid l \in \Lcov{k}{P} \}$$
		If $\Hyp$ and $\Mod$ are $Z$-equivalent, then $\bisim$ is a bisimulation relation.
		\label{th:Z_bisim}
	\end{theorem}
	
	\begin{corollary}
		$\Hyp$ and $\Mod$ are $Z$-equivalent if and only if they are equivalent.
		\label{cor:Z_exhaustive}
	\end{corollary}
	
	We can infer a state cover $P$ and a characterization set $W$ from $\PLlambda$'s $S$ and $\C$;
	thus, as long as parameter $k$ is known, we can substitute EQs with MQs over $Z$.	
	Counting terms in $\Lcov{k+1}{P}$ results in the following bound:
	\begin{theorem}
		\label{th:size_Z}
		$|Z| \leq n \cdot (|\Sigma| + n)^{2^{k+1}}$.
	\end{theorem}

	\section{Experimental Results}
	\label{sec:exp}
	
	We implemented $\PLlambda$ and \FindEBP as well as van Heerdt et al.'s $\PLstar$~\cite{DBLP:conf/fossacs/HeerdtKR021} and \HCE in a \texttt{C++} prototype\footnotemark.
	\footnotetext{\texttt{https://gitlab.lre.epita.fr/adrien/treelearn}}
	%\footnotetext{We omit a link to preserve anonymity, but plan on submitting an artifact later.}
	Since \FindEBP and \HCE can be used by $\PLlambda$ and $\PLstar$ alike, we ran different experiments in order to independently assess how the various learning algorithms and counterexample analysis procedures impact query and symbol complexity.
	To evaluate the impact of arbitrary long counterexamples, minimal counterexamples returned by the MAT were artificially lengthened by pumping loops in the product PR $\Hyp \oplus \Mod$ (built in a similar fashion to a product tree automaton~\cite{comon2008tree}).
	
	We benchmarked the four algorithm combinations against a sample of $236$ randomly generated PRs, as shown in Figure~\ref{fig:logplot_bench}.
	Points above the diagonal line correspond to experiments where $\PLlambda$ and \FindEBP outperformed another combination of options.
	Table \ref{fig:table_bench} displays a summary of these benchmarks.
	
	{\centering
		\begin{minipage}{0.46\linewidth}
			\begin{figure}[H]
				\setlength{\tabcolsep}{5pt}
				\renewcommand{\arraystretch}{1.25}
				\scalebox{0.9}{\begin{tabular}{|c|ccc|} 
						\hline
						& \textbf{Mem.} & \textbf{Eq.} & \textbf{Symb.} \\
						\hline
						$\PLlambda$ + \FindEBP & 296.17 & 2.93 & 5763.8 \\ 
						$\PLlambda$ + \HCE & 325.49 & 2.92 & 9594.1 \\
						$\PLstar$ + \FindEBP & 486.38 & 2.73 & 30911 \\
						$\PLstar$ + \HCE & 515.05 & 2.70 & 40943 \\
						\hline
				\end{tabular}} 
				\caption{Summary of the arithmetic mean of the \textbf{Mem}bership, \textbf{Symb}ol, and \textbf{Eq}uivalence complexities.}
				\label{fig:table_bench}
			\end{figure}
		\end{minipage}
		\hfill
		\begin{minipage}{0.5\linewidth}
			Algorithm $\PLlambda$ (resp. \FindEBP) on average outperforms $\PLstar$ (resp. \HCE) in terms of membership and symbol complexity.
			In particular, the combination of $\PLlambda$ and \FindEBP outperforms $\PLstar$ and \HCE in every single instance save two, and the latter's membership (resp. symbol) complexity is on average 1.74 (resp. 7.10) times higher.
			However, $\PLlambda$ and \FindEBP's equivalence complexity is on average 1.08 times higher, a meagre difference explained by compatibility tests.
	\end{minipage}}

	The introduction of $\PLlambda$ yields the greatest benefits.
	Nevertheless, the use of \FindEBP instead of \HCE also results in a significant symbol complexity improvement—likely due to entire subterms of the counterexample being pruned by \FindEBP—as well as more modest membership complexity gains.
	
	Assuming the learning algorithm used is $\PLlambda$, it's worth noting that \HCE yields marginal membership complexity gains in 28\% of all cases, a possible explanation being that \HCE performs bottom-up substitutions by access pomsets while \FindEBP prunes branches in a top-down fashion;
	thus, if a refinement can be directly inferred from the leftmost leaf, \HCE will outperform \FindEBP.
	Nevertheless, \FindEBP's theoretical complexity guarantees a significantly better worst case scenario.
	
	\begin{figure}[H]
		\centering
		\resizebox{!}{4.8cm}{\input{bench_table_mem.tex}}
		\hfill
		\resizebox{!}{4.8cm}{\input{bench_table_symb.tex}}
		\caption{Comparing $\PLlambda$ to $\PLstar$ and \FindEBP to \HCE.}
		\label{fig:logplot_bench}
	\end{figure}

	\section{Conclusion and Further Developments}	
	
	Our active learning algorithm $\PLlambda$ and our new counterexample analysis algorithm \FindEBP have been empirically shown to yield significant membership and symbol complexity gains.
	But the following practical issues we successfully tackled or are currently dealing with are also worth mentioning:
	\begin{description}
		\item[Representing series-parallel pomsets.] We settled on a \emph{canonical} representation of pomsets as binary trees that guarantees minimal depth and prevents duplicate queries.
		
		\item[Generating a benchmark.] We used a SAT-based approach to create a test sample of minimal, reachable pomset recognizers.
		
		\item[Benchmarking the $W$-method.] Our $W$-method has yet to be implemented.
		It remains to be seen whether its high cost dominates the learning process.
		We plan on adapting various improvements to the $W$-method, such as the $H$-method~\cite{dorofeeva05hmethod}, to PRs.
		Random walks are also a lead worth exploring.
				
		\item[Optimizing counterexample handling.] \texttt{FindEBP} benefits from the use of canonical terms of minimal depth.
		However, its complexity remains linear w.r.t. the input's size if the canonical term is a linear tree.
		We may therefore try to develop another counterexample handling procedure optimized for linear trees, and even seek one that is truly logarithmic.
	\end{description}
	
	Another lead consists in determining whether \texttt{FindEBP} and $\PLlambda$ can enhance AL for general (non-binary) tree languages~\cite{DBLP:conf/dlt/DrewesH03}.
	We also plan on exploring the \emph{passive learning} problem for pomset samples:
	given two non-empty sets $Z^+ \subseteq \SP$ and $Z^- \subseteq \SP$ such that $Z^+ \cap Z^- = \emptyset$, find a PR that accepts all the elements of $Z^+$ and rejects all the elements of $Z^-$.
	
	Moreover, some use cases such as producer-consumer systems require modelling pomsets that feature $N$ patterns.
	These scenarios can be effectively formalized using \emph{interval pomsets}~\cite{DBLP:journals/iandc/FahrenbergJSZ22} (that is, pomsets such that $x_1 < x_2$ and $x_3 < x_4$ implies $x_3 < x_2$ or $x_1 < x_4$) recognized by \emph{higher-dimensional automata} (HDA)~\cite{DBLP:conf/concur/FahrenbergJSZ22}.
	The extension of Myhill-Nerode's theorem to languages of HDA~\cite{DBLP:journals/fuin/FahrenbergZ24} opens up the possibility of an AL algorithm for HDA.
	The class of HDA supplements rather than subsumes PRs:
	indeed, $N$-shaped pomsets are interval but not series-parallel, while pomset $ab \parallel cd$ is series-parallel but not interval. 
	
	Finally, designing a practical oracle capable of processing pomset-shaped queries is a significant technical challenge:
	submitting linearized (as mentioned in Section~\ref{sec:related}) pomset queries to an ordinary sequential oracle is not a suitable solution as it may cause the algorithm's membership complexity to blow up.
	
	\paragraph{Data availability statement.}
	The data that support the findings of this article are openly available at: 
	\texttt{https://doi.org/10.5281/zenodo.18175343}

	\bibliographystyle{plain} 
	\bibliography{mybib}

	\newpage
	\appendix
	
	\section{Properties of the Hypothesis}
	\label{app:properties}
	
	$\Llambda$ \cite{DBLP:conf/birthday/HowarS22} on finite automata maintains a prefix-closed set of access sequences and a suffix-closed set of distinguishing suffixes.
	We show that similar results hold for $\PLlambda$ as well.
	
	\begin{property}[Closedness of access pomsets]
		\label{prop:s_subclosed}
		$S$ is subpomset-closed.
	\end{property}
	
	\begin{proof}
		With the exception of $\varepsilon$, new elements of $S$ are only ever added by promoting elements of $S^+$ (Line~\ref{algo:expand:init_end} of Algorithm~\ref{algo:expand}, Lines~\ref{algo:refine:b0_no_as} and \ref{algo:refine:b1_no_as} of Algorithm~\ref{algo:refine}, Line~\ref{algo:analyze_ce:new_component} of Algorithm~\ref{algo:analyze_ce});
		$S^+$ itself is inductively defined as the union of $\Sigma$ and the composition of elements of $S$. \hfill\qed
	\end{proof}
	
	Similarly, $S \cup S^+$ is subpomset-closed.
	We show below that an associative, consistent, and compatible $(\B, S, \D)$ induces a minimal hypothesis $\Hyp$.
	
	\begin{lemma}[Reachability of $\Hyp$]
		\label{lem:reach}
		An associative, consistent $(\B, S, \D)$ induces a hypothesis $\Hyp$ such that:
		\begin{enumerate*}
			\item $\eval_\Hyp(s) = \B_s$ for all $s \in S \cup S^+$,
			\item $\Hyp$ is reachable.
		\end{enumerate*}
	\end{lemma}
	
	\begin{proof}
		$S \cup S^+$ being subpomset closed, we can rely on a proof by induction on $s$ to prove that for any $s \in S \cup S^+$, $\eval_\Hyp(s) = \B_s$.
		\begin{description}
			\item[Base case.] $\eval_H(\varepsilon) = i_\Hyp(\varepsilon) = \neut_\Hyp = \B_\varepsilon$ and, for any $a \in \Sigma$, $\eval_\Hyp(a) = i_\Hyp(a) = \B_a$ by definition of the hypothesis $\Hyp$.
			
			\item[Inductive case.] 	Now, let $s_1 \circ s_2 \in S \cup S^+$ for some $\circ \in \operators$;
			$S \cup S^+$ being subpomset-closed, $s_1, s_2 \in S \cup S^+$.
			By induction hypothesis, $\eval_H(s_i) = \B_{s_i}$ for $i \in \{1, 2\}$.
			Then $\eval_\Hyp(s_1 \circ s_2) = \eval_\Hyp(s_1) \circ_\Hyp \eval_\Hyp(s_2) = \B_{s_1} \circ_\Hyp \B_{s_2} = \B_{s_1 \circ s_2}$ by definition of $\Hyp$ and consistency of $\B$. Thus \textbf{1.} holds.
		\end{description}
		\textbf{2.} is a direct consequence of $\eval_\Hyp(s) = \B_s$ for any $s \in S$. \hfill\qed
	\end{proof}
	
	\begin{lemma}[Partial compatibility]
		\label{lem:s_splus_compatibility}
		An associative, consistent $(\B, S, \D)$ induces a $(S \cup S^+)$-compatible hypothesis $\Hyp$.
	\end{lemma}
	
	\begin{proof}
		By Lemma~\ref{lem:reach}, for any $s \in S \cup S^+$, $\eval_\Hyp(s) = \B_s$.
		By definition of $\Hyp$, $\Hyp(s) = 1 \iff \eval_\Hyp(s) \in F_\Hyp \iff \B_s \in F_\Hyp \iff \forall s' \in \B_s, \Mod(s') = 1$.
		Thus, $\Hyp(s) = \Mod(s)$. \hfill\qed
	\end{proof}
	
	\begin{property}[Closedness of distinguishing contexts]
		\label{prop:c_subclosed}
		$\C$ is such that, for any $c \in \C$, either $c = \square$ or there exist $c' \in \C$, $s \in S \setminus \{\varepsilon\}$, and $\circ \in \operators$ such that $c = c'[\square \circ s]$ or $c = c'[s \circ \square]$.
	\end{property}
	
	\begin{proof}
		With the exception of $\square$, new elements of $\C$ are only ever added by Line~\ref{algo:make_consistent:refine} of Algorithm~\ref{algo:make_consistent} or Lines~\ref{algo:make_assoc:refine_1} and \ref{algo:make_assoc:refine_2} of Algorithm~\ref{algo:make_assoc};
		both additions follow the desired pattern. \hfill\qed
	\end{proof}
	
	\begin{theorem}[Minimality]
		\label{th:minimal}
		Given a hypothesis $\Hyp$ induced from an associative, consistent $(\B, S, \D)$, if $\Hyp$ is compatible, then it is minimal.
	\end{theorem}
	
	\begin{proof}
		First, by Lemma~\ref{lem:reach}, we know that $\Hyp$ is reachable.
		Let $B_1, B_2 \in \B$ be such that $B_1 \neq B_2$.
		By design of the DT and of the partition $\B$, there exist $c \in \C_{B_1} \cap \C_{B_2}$, $s_1 \in B_1$, $s_2 \in B_2$, $\Mod(c[s_1]) \neq \Mod(c[s_2])$.
		$\Hyp$ being compatible, $\Hyp(c[s_1]) = \Mod(c[s_1])$ and $\Hyp(c[s_2]) = \Mod(c[s_2])$.
		Thus, $\Hyp(c[s_1]) \neq \Hyp(c[s_2])$.
		$\Hyp$ is therefore minimal. \hfill\qed
	\end{proof}

	\section{Termination and Correctness of \FindEBP}
	\label{app:proof_find_ebp}
	
	We assume in this section that $\B$ is a consistent, associative partition from which a hypothesis $\Hyp$ is inferred.
	
	The following theorem is one of our main results:
	it states that from a counterexample, by looking for breaking points, we can find a representative $p$ of a new component that is distinguished by a context $c'$ from any other representative $p'$ of its current component.
	
	\begin{theorem}[Correction and termination of Algorithm \ref{algo:find_ebp}]
		\label{th:find_ebp}
		Given $c \in \Co$ and a minimal $t \in \ST{z}$ such that $\Hyp(c[z]) \neq \Mod(c[z])$ and $\agree(c, z) = 1$, $\mathtt{FindEBP}(c, t)$ terminates and returns a pair $(c', p) \in \Co \times S^+$ such that $\forall p' \in \as_\Hyp(p)$, $\Mod(c'[p]) \neq \Mod(c'[p'])$.
	\end{theorem}
	
	\begin{proof}
		We prove this theorem by induction on the depth $\depth{z}$ of term $t$.
		\begin{description}
			\item[Base case.] If $\depth{z} = 0$, $z \in \Sigma$ and $z \in S \cup S^+$ by definition of $S^+$.
			
			Let us prove that $z \in S^+$
			Due to $\agree(c, z) = 1$, for any $p' \in \as_\Hyp(z)$, $\Hyp(c[p']) = \Mod(c[p'])$.
			But $\Hyp(c[z]) \neq \Mod(c[z])$ by hypothesis;
			as a direct consequence, $z \not\in \as_\Hyp(z)$, hence $z \not\in S$, thus $z \in S^+$.
			
			Moreover, $\Mod(c[z]) \neq \Mod(c[p'])$.
			
			The base case therefore holds for $c' = c$ and $p = z$.
			
			\item[Inductive case.] Assume that $\depth{z} = m+1$ for some $m \in \integers$ and that the theorem holds for smaller depths.
			Let $t = t_1 \circ t_2$, where $t_i \in \ST{z_i}$ for $i \in \{1, 2\}$.
			
			We perform a case disjunction on $\agree(c[\square \circ z_2], t_1)$:
			\begin{description}
				\item[It is true.] Since $c[\square \circ z_2][z_1] = c[z]$ and $\Hyp(c[z]) \neq \Mod(c[z])$, the preconditions of Line~\ref{algo:find_ebp:find_bp_end} hold and $\mathtt{FindEBP}(c, t) = \mathtt{FindEBP}(c[\square \circ z_2], t_1)$.
				
				Since $\depth{z_1} < \depth{z}$ by minimality of $t$, we can apply the induction hypothesis to $\mathtt{FindEBP}(c[\square \circ z_2], t_1)$ and the desired result holds.
				
				\item[It is false.] Then $\exists p_1 \in \as_\Hyp(z_1)$, $\Hyp(c[p_1 \circ z_2]) \neq \Mod(c[p_1 \circ z_2])$ \textbf{(\romannumeral 1)}.
				We perform a case disjunction on predicate $\agree(c[p_1 \circ \square], z_2)$.
				\begin{description}
					\item[It is true.] 
					By \textbf{\romannumeral 1}, $c[p_1 \circ \square][z_2]$ is a counterexample.
					The preconditions of Line~\ref{algo:find_ebp:look_for_ebp} thus hold and $\mathtt{FindEBP}(c, t) = \mathtt{FindEBP}(c[p_1 \circ \square], t_2)$.
					
					Since $\depth{z_2} < \depth{z}$ due to $t$ being minimal, we can again apply the induction hypothesis and the desired result holds. 
					
					\item[It is false.] By definition of the agreement predicate, $\exists p_2 \in \as_\Hyp(z_2)$ such that $\Hyp(c[p_1 \circ p_2]) \neq \Mod(c[p_1 \circ p_2])$ \textbf{(\romannumeral 2)}.
					
					Consider $p = p_1 \circ p_2 \in S \cup S^+$.					
					By definition, $\Hyp(z) = \Hyp(z_1 \circ z_2)$, and by Property \ref{prop:freeness_as}, $\Hyp(z_1 \circ z_2) = \Hyp(p_1 \circ p_2) = \Hyp(c[p])$.
					Thus, $p \in \B_z$.
					
					Let us prove that $p \in S^+$.
					By hypothesis, since $\agree(c, z) = 1$, for any $p' \in \as_\Hyp(z)$, $\Hyp(c[p']) = \Mod(c[p'])$ must hold \textbf{(\romannumeral 3)}.
					Therefore, by \textbf{\romannumeral 2}, $p \not\in \as_\Hyp(z)$, thus $p \not\in S$.
					As a consequence, $p \in S^+$.
					
					By Property~\ref{prop:freeness_as}, $ \Hyp(c[p]) = \Hyp(c[p'])$ for any $p' \in \as_\Hyp(z)$.
					By \textbf{\romannumeral 2}, $\Mod(c[p]) \neq \Hyp(c[p])$, and by \textbf{\romannumeral 3}, $\Mod(c[p']) \neq \Mod(c[p])$.
					
					Finally, the theorem holds for $c' = c$ and $p$. \hfill\qed		
				\end{description}
			\end{description}
		\end{description}
	\end{proof}
	
	Finally, \FindEBP results in a refinement of our under-approximation of $\sim_L$:
	
	\begin{corollary}
		\label{cor:new_component}
		Given a counterexample $w \in \SP$ such that $\Hyp(w) \neq \Mod(w)$, a call to $\mathtt{FindEBP}(\square, w)$ returns a pair $(c', p) \in \Co \times S^+$ such that for any $p' \in S$, $p \not\sim_L p'$.
	\end{corollary}
	
	\begin{proof} 
		By Lemma~\ref{lem:s_splus_compatibility}, $\forall p \in \as_\Hyp(w)$, $\Hyp(p) = \Mod(p)$.
		Thus, $\agree(\square, w) = 1$.
		We can therefore apply Algorithm~\ref{algo:find_ebp} to $(\square, w)$ and by Theorem~\ref{th:find_ebp}, $\mathtt{FindEBP}(\square, w)$ returns a pair $(c, p) \in \Co \times S^+$ such that for any $p' \in \as_\Hyp(p)$, $\Mod(c[p]) \neq \Mod(c[p'])$.
		Moreover, by design of DT $\D$ and partition $\B$, for any $p' \in S \setminus \as_\Hyp(p)$, there exists $c' \in \C$ such that $\Mod(c'[p]) \neq \Mod(c'[p'])$. 
	\end{proof}

	\section{Generating Test Suites for Equivalence Queries}
	
	The definitions of minimality and characterization sets trivially result in the following property:
	
	\begin{property}
		\label{prop:W_equiv}
		Given a minimal pomset recognizer $\PR = (R, \odot, \obar, \neut, i, F)$, a characterization set $W$ of $P$, and two states $r_1, r_2 \in R$, if for any $c \in W, \PR(c[r_1]) = \PR(c[r_2])$, then $r_1 = r_2$.
	\end{property}
		
	\begin{center}
		\begin{minipage}{0.67\linewidth}
			In order to prove Theorem~\ref{th:canonical_state_cover}, we introduce the \emph{evaluation tree} (ET) $e_\PR(t)$ of a term $t \in \ST{w}$ w.r.t. a PR $\PR = (M, \odot, \obar, \neut, i, F)$.
			It is a relabelling of $t$ defined inductively as follows:
			\begin{itemize}
				\item $e_\PR(a) = i(a)$ for $a \in \Sigma$;
				
				\item $e_\PR(t_1 \circ t_2) = \eval(e_\PR(t_1) \circ e_\PR(t_2))$.
			\end{itemize}
			Tree $e_\PR(t)$ is also said to be an ET of pomset $w$ w.r.t. $\PR$.
			Figure~\ref{fig:eval_tree} displays an ET that explicits the bottom-up computation performed by PR $\PR$ of Example~\ref{ex:PR} on pomset $bc \parallel a$.
		\end{minipage}
		\hfill
		\begin{minipage}{0.3\linewidth}
			\begin{figure}[H]
				\centering
				\begin{tikzpicture}[sibling distance=1.5cm, level distance=1cm]
					\node {$r_c$}
					child {node (l) {$r_{bc}$}
						child {node (ll) {$r_b$}}
						child {node (lr) {$r_c$}}}
					child {node {$r_a$}};
				\end{tikzpicture}
				\caption{An evaluation tree of $bc \parallel a$.}
				\label{fig:eval_tree}
			\end{figure}
		\end{minipage}
	\end{center}

	\subsection{Proof of Theorem~\ref{th:canonical_state_cover}}
	\label{app:proof_state_cover}
	
	\begin{proof}
		We first prove that $P$ covers at least $n$ distinguishable states of $\Mod$ \textbf{(\romannumeral 1)}.
		
		Indeed, consider two distinguishable states $h_1$ and $h_2$ among the $n$ states of $\Hyp$.
		Let $p_1$ (resp. $p_2$) be an access pomset of $h_1$ (resp. $h_2$) in $P$.
		$W$ being a characterization set of $\Hyp$, $\exists c \in W$, $\Hyp(c[p_1]) \neq \Hyp(c[p_2])$.
		But $\Hyp \equiv_{W[P]} \Mod$ and $c[p_1], c[p_2] \in W[P]$, therefore $\Mod(c[p_1]) \neq \Mod(c[p_2])$;
		states $\eval_\Mod(p_1)$, $\eval_\Mod(p_2)$ are distinguished in $\Mod$ and \textbf{\romannumeral 1} holds.
		
		Consider a state $q \in M$ and an access pomset $w$ of $q$ in $\Mod$.
		Let us prove that there exists an access pomset $w'$ of $q$ and an evaluation tree $\tau'$ of a term $t \in \ST{w}$ w.r.t. $\Mod$ that admits no repetition of states along any branch \textbf{(\romannumeral 2)}.
		Assume that such a repetition occurs in an evaluation tree $\tau$ of $w$, as shown in Figure~\ref{fig:eval_w}:
		then there exist two contexts $c_1, c_2 \in \Co$, a pomset $z \in \SP$, and a state $q' \in M$ such that $w = c_1[c_2[z]]$ and $q' = \eval_\Mod(z) = \eval_\Mod(c_2[z])$.
		
		Let $w' = c_1[z]$.
		Then $\eval_\Mod(w') = \eval_\Mod(c_1[z]) = \eval_\Mod(c_1[c_2[z]]) = \eval_\Mod(w)$, and we can remove at least one repetition from the branches of $\tau$, as shown in Figure~\ref{fig:eval_w_p}.
		Repeating this procedure yields a term with an execution tree that features no repetitions at all along its branches and evaluates to $q$;
		\textbf{\romannumeral 2} thus holds.
		
		\begin{figure}
			\begin{minipage}{0.48\linewidth}
				\begin{figure}[H]
					\centering
					\begin{tikzpicture}[sibling distance=15mm]
						\node (root) {$q$} [level distance=12mm]
						child {node[blue] (l) {\fbox{$\textcolor{red}{q'}$}} edge from parent[-, dashed]
							child {node[blue] (ll) {\fbox{$\textcolor{red}{q'}$}} [level distance=9mm]
								child {node (llbelow) {\ldots} edge from parent[draw=none]}}
							child {node (lr) {\ldots}}}
						child {node (r) {\ldots} edge from parent[-, dashed]};
						\node[draw=blue, dotted, label={0:$\textcolor{blue}{\mathbf{c_1}}$}, fit=(root) (l) (r)] {};
						\node[draw=blue, dotted, label={180:$\textcolor{blue}{\mathbf{c_2}}$}, fit=(l) (ll) (lr)] {};
						\node[draw=purple, trapezium, dotted, label={270:$\textcolor{purple}{\mathbf{z}}$}, fit=(ll) (llbelow)] {};
					\end{tikzpicture}
					\caption{An evaluation tree $\tau$ of $w$.}
					\label{fig:eval_w}
				\end{figure}
			\end{minipage}
			\hfill
			\begin{minipage}{0.48\linewidth}
				\begin{figure}[H]
					\centering
					\begin{tikzpicture}[sibling distance=15mm]
						\node (root) {$q$} [level distance=12mm]
						child {node[blue] (l) {\fbox{$\textcolor{red}{q'}$}} edge from parent[-, dashed, level distance=9mm]
							child {node (lbelow) {\ldots} edge from parent[draw=none]}}
						child {node (r) {\ldots} edge from parent[-, dashed]};
						\node[draw=blue, dotted, label={0:$\textcolor{blue}{\mathbf{c_1}}$}, fit=(root) (l) (r)] {};
						\node[draw=purple, trapezium, dotted, label={270:$\textcolor{purple}{\mathbf{z}}$}, fit=(l) (lbelow)] {};
					\end{tikzpicture}
					\caption{A factorized tree $\tau'$ of $w'$.}
					\label{fig:eval_w_p}
				\end{figure}
			\end{minipage}
		\end{figure}
		
		By \textbf{\romannumeral 2}, consider an access pomset $w' \in \SP$ of $q$ that admits an ET $\tau'$ without repetitions.
		If a branch of $\tau'$ is of length at least $k+1$, by applying the pigeonhole principle, then at least one of the $k+1$ first states occurring from the root along this branch is covered by $P$, due to $\Mod$ having $n+k$ states and $P$ covering at least $n$ states by \textbf{\romannumeral 1}, leaving at most $k$ states uncovered by $P$.
		
		More formally, as displayed in Figure \ref{fig:w_p_uncovered}, there exist $m \in \integers$, a multi-context $c \in \Ct{m}$, and $z_1, \ldots, z_m \in \SP$ such that $\delta(c) \leq k$, $w' = c[z_1, \ldots, z_m]$ and for any $j \in \{1, \ldots, m\}$, state $\eval_\Mod(z_j) = q_j$ is covered by $P$.
		
		Let $p_j \in P$ be an access pomset of $q_j$ in $\Mod$, that is, $\eval_\Mod(p_j) = q_j$.
		Then consider $p = c[p_1, \ldots, p_m]$, as shown in Figure \ref{fig:w_p_covered}.
		By design, $p \in \Lcov{k}{P}$.
		Moreover, $\eval_\Mod(p) = \eval_\Mod(c[p_1, \ldots, p_m]) = \eval_\Mod(c[z_1, \ldots, z_m]) = \eval_\Mod(w') = q$.
		
		Thus, every state of $\Mod$ is covered by $\Lcov{k}{P}$, and this set is therefore a state cover of $\Mod$. \hfill\qed
	\end{proof}
	
	\begin{figure}
		\begin{minipage}{0.48\linewidth}
			\begin{figure}[H]
				\begin{tikzpicture}[sibling distance=32mm]
					\node (root) {$q$} [level distance=12mm]
					child {node[blue] (l) {\fbox{$\textcolor{black}{q_1}$}} edge from parent[-, dashed, level distance=9mm]
						child {node (lbelow) {\ldots} edge from parent[draw=none]}}
					child {node[blue] (r) {\fbox{$\textcolor{black}{q_2}$}} edge from parent[-, dashed, level distance=9mm]
						child {node (rbelow) {\ldots} edge from parent[draw=none]}};
					\node[draw=blue, dotted, label={0:$\textcolor{blue}{\mathbf{c}}$}, fit=(root) (l) (r)] {};
					\node[draw=purple, trapezium, dotted, label={270:$\textcolor{purple}{\mathbf{z_1}}$}, fit=(l) (lbelow)] {};
					\node[draw=purple, trapezium, dotted, label={270:$\textcolor{purple}{\mathbf{z_2}}$}, fit=(r) (rbelow)] {};
				\end{tikzpicture}
				\caption{Finding states covered by $P$ in an ET $\tau'$ of $w'$.}
				\label{fig:w_p_uncovered}
			\end{figure}
		\end{minipage}
		\hfill
		\begin{minipage}{0.48\linewidth}
			\begin{figure}[H]
				\begin{tikzpicture}[sibling distance=32mm]
					\node (root) {$q$} [level distance=12mm]
					child {node[blue] (l) {\fbox{$\textcolor{black}{q_1}$}} edge from parent[-, dashed, level distance=9mm]
						child {node (lbelow) {\ldots} edge from parent[draw=none]}}
					child {node[blue] (r) {\fbox{$\textcolor{black}{q_2}$}} edge from parent[-, dashed, level distance=9mm]
						child {node (rbelow) {\ldots} edge from parent[draw=none]}};
					\node[draw=blue, dotted, label={0:$\textcolor{blue}{\mathbf{c}}$}, fit=(root) (l) (r)] {};
					\node[draw=red, trapezium, dotted, label={270:$\textcolor{red}{\mathbf{p_1}}$}, fit=(l) (lbelow)] {};
					\node[draw=red, trapezium, dotted, label={270:$\textcolor{red}{\mathbf{p_2}}$}, fit=(r) (rbelow)] {};
				\end{tikzpicture}
				\caption{Inserting $P$'s access pomsets in $\tau'$.}
				\label{fig:w_p_covered}
			\end{figure}
		\end{minipage}
	\end{figure}

	\subsection{Proof of Theorem~\ref{th:Z_bisim}}
	\label{app:proof_bisim}
	
	We will use a proof inspired by Moerman's~\cite{Moerman19} that relies on bisimulation to prove Theorem~\ref{th:Z_bisim}.
	
	\begin{proof}
		Note that $W[P] \subseteq Z$.
		$\Hyp$ and $\Mod$ being $Z$-equivalent, they are also $W[P]$-equivalent, and by Theorem \ref{th:canonical_state_cover}, $\Lcov{k}{P}$ is a state cover of $\Mod$. Let us prove that ${\bisim}$ is a bisimulation relation.
		\begin{enumerate}
			\item Let $(h, m) = (\eval_\Hyp(l), \eval_\Mod(l)) \in {\bisim}$.
			Note that $\square \in W$ and $l \in \Lcov{k}{P}$, thus $l = \square[l] \in Z$ and $\Hyp(l) = \Mod(l)$.
			Hence, $h \in F_\Hyp \iff m \in F_\Mod$.
			
			\item Let $(h_1, m_1) = (\eval_\Hyp(l_1), \eval_\Mod(l_1)) \in {\bisim}$, $(h_2, m_2) = (\eval_\Hyp(l_2), \eval_\Mod(l_2)) \in {\bisim}$, and $\circ \in \operators$.
			Let us prove that $h_1 \circ_\Hyp h_2 \bisim m_1 \circ_\Mod m_2$.
			
			We consider an access pomset $v$ of $l_1 \circ_\Mod l_2$ in the state cover $\Lcov{k}{P}$ of $\Mod$;
			then $\eval_\Mod(l_1 \circ \l_2) = \eval_\Mod(v)$ \textbf{(\romannumeral 1)}.
			Thus, $\forall c \in W$, $\Mod(c[l_1 \circ l_2]) = \Mod(c[v])$ \textbf{(\romannumeral 2)}.
			
			Note that $W[\{v\}] \subseteq Z$, thus for any $c \in W$, $\Hyp(c[v]) = \Mod(c[v])$\textbf{(\romannumeral 3)}.
			
			Moreover, $l_1 \circ l_2 \in \Lcov{k+1}{P}$ due to $l_1, l_2 \in \Lcov{k}{P}$.
			Thus, $W[\{l_1 \circ \l_2\}] \subseteq Z$ and for any $c \in W$, $\Hyp(c[l_1 \circ l_2]) = \Mod(c[l_1 \circ l_2])$.
			By \textbf{\romannumeral 2}, $\Hyp(c[l_1 \circ l_2]) = \Mod(c[v])$ holds \textbf{(\romannumeral 4)}.
			
			By \textbf{\romannumeral 3} and \textbf{\romannumeral 4}, $\forall c \in W$, $\Hyp(c[v]) = \Hyp(c[l_1 \circ l_2])$.
			But $W$ is a characterization set of $\Hyp$, therefore $\eval_\Hyp(l_1 \circ l_2) = \eval_\Hyp(v)$ \textbf{(\romannumeral 5)} by Property \ref{prop:W_equiv}.
			
			By definition of $\bisim$, $\eval_\Hyp(v) \bisim \eval_\Mod(v)$.
			Thus, by \textbf{\romannumeral 1} and \textbf{\romannumeral 5}, $\eval_\Hyp(l_1 \circ l_2) \bisim \eval_\Mod(l_1 \circ l_2)$.
			This result is not trivial due to $l_1 \circ l_2$ being an element of $\Lcov{k+1}{P}$ but possibly not of $\Lcov{k}{P}$, hence being excluded from $\sim$'s definition.
			
			We have thus proven $h_1 \circ_\Hyp h_2 \bisim m_1 \circ_\Mod m_2$.
		\end{enumerate}
		${\bisim}$ is therefore a bisimulation relation.
	\end{proof}
	
	This result yields a proof of Corollary~\ref{cor:Z_exhaustive}.
	
	\begin{proof}
		If $\Hyp$ and $\Mod$ are $Z$-equivalent, $\bisim$ is a bisimulation relation by Theorem \ref{th:Z_bisim}.
		Moreover, by definition of $\bisim$, $\forall x \in \Sigma \cup \{\varepsilon\}$, obviously $\eval_\Hyp(x) \bisim \eval_\Mod(x)$.
		By Property \ref{prop:bisim_equiv}, $\Hyp$ and $\Mod$ are equivalent.
		The converse is trivial.
	\end{proof}

	\subsection{Proof of Theorem~\ref{th:size_Z}}
	\label{app:proof_size_Z}	
	
	\begin{proof}
		Trivially, $|W| \leq n$ due to $W$ being minimal:
		it takes at most $n$ contexts to distinguish $n$ states.
		Similarly, $|P| \leq n$ due to $\Hyp$ having $n$ states.
		Let $T_P$ (resp. $T_W$) be a set of $|P|$ (resp. $|W|$) terms such that each pomset in $P$ (resp. $W$) admits exactly one representative in $T_P$ (resp. $T_W$).
		We define:
		\begin{itemize}
			\item The set $U_{k+1}$ of $\varepsilon$-free terms of depth lesser than or equal to $k+1$ representing multi-contexts over the alphabet $\Sigma$.
			
			\item The set $V_{k+1}$ of terms over $\Sigma$ obtained by inserting terms belonging to $T_P$ in a context belonging to $U_{k+1}$.
			
			\item The set $X_{k+1}$ of terms over $\Sigma$ obtained by inserting a term belonging to $V_{k+1}$ in a context belonging to $T_W$.
		\end{itemize}
		
		The following properties then hold:
		\begin{itemize}
			\item Trivially, every pomset in $Z$ admits a representative in $X_{k+1}$.
			Since pomsets are equivalence classes over terms, $|Z| \leq |X_{k+1}|$.
			
			\item We can interpret $V_{k+1}$ as the set of binary trees of depth lesser than or equal to $k+1$ labelled by $\Sigma \cup P$ (leaves) and $\operators$ (internal nodes).
			And there are less than $|\Sigma| + |P|)^{2^{k+1}}$ such trees.
			Thus, $|V_{k+1}| \leq (|\Sigma| + n)^{2^{k+1}}$.
			
			\item Obviously, $|X_{k+1}| \leq |T_W| \cdot |V_{k+1}|$.
		\end{itemize}
		
		We therefore conclude that $|Z| \leq n \cdot (|\Sigma| + n)^{2^{k+1}}$. \hfill\qed
	\end{proof}
	
	It is possible to determine whether $\Hyp$ and $\Mod$ are equivalent by performing an indirect emptyness check of the product PR $\Mod \bigoplus \Hyp$ that consists in querying every pomset of depth smaller than or equal to $n \cdot (n + k)$.
	However, this naive algorithm requires at most $|\Sigma|^{2^{n \cdot (n + k)}}$ MQs \textbf{(\romannumeral 1)}, and thus happens to be more expensive than the $W$-method.
	
	In particular, the most expensive simulated EQs are the ones that yield a positive answer:
	the full sample has to be queried rather than halting the process as soon as a counterexample has been found.
	But in these cases, $k$ is at its lowest value of the entire learning process, and Theorem~\ref{th:size_Z} provides a tighter upper bound than \textbf{\romannumeral 1}.

\end{document}